\documentclass[fleqn,usenatbib,useAMS]{mnras}

\usepackage{graphicx}	
\usepackage{amsmath}	
\usepackage{amssymb}	
\usepackage{multicol}        
\usepackage{bm}		
\usepackage{pdflscape}	
\usepackage[normalem]{ulem}	
\usepackage[usenames]{color}
\graphicspath{{./old_figs/}{./}}

\newcommand{\beq}{\begin{equation}}
\newcommand{\eeq}{\end{equation}}
\newcommand{\beqa}{\begin{eqnarray}}
\newcommand{\eeqa}{\end{eqnarray}}
\newcommand{\nn}{\nonumber}
\newcommand{\mbf}{\bm}
\newcommand{\grad}{\nabla}
\newcommand{\vgrad}{\bm{\nabla}}

\usepackage[T1]{fontenc}
\usepackage{ae,aecompl}

\usepackage{newtxtext,newtxmath}

\title[I-Love-Q Relations for Realistic White Dwarfs]{I-Love-Q Relations for Realistic White Dwarfs}

\author[Andrew Taylor, Kent Yagi, Phil L. Arras]{Andrew J. Taylor$^{1}$\thanks{Contact e-mail: \href{mailto:ajt9gx@virginia.edu}{ajt9gx@virginia.edu}}, Kent Yagi$^{2}$\thanks{Contact e-mail: \href{mailto:ky5t@virginia.edu}{ky5t@virginia.edu}}, Phil L. Arras$^{1}$\thanks{Contact e-mail: \href{mailto:arras@virginia.edu}{arras@virginia.edu}}
\\
$^{1}$Department of Astronomy, University of Virginia, Charlottesville, VA\\
$^{2}$Departmnet of Physics, University of Virginia, Charlottesville, VA}

\date{Accepted XXX. Received YYY; in original form ZZZ}

\pubyear{2019}

\begin{document}
\label{firstpage}
\pagerange{\pageref{firstpage}--\pageref{lastpage}}
\maketitle

\begin{abstract}
The space-borne gravitational wave interferometer, LISA, is expected to detect signals from numerous binary white dwarfs. At small orbital separation, rapid rotation and large tidal bulges may allow for the stellar internal structure to be probed through such observations.
Finite-size effects are encoded in quantities like the moment of inertia ($I$), tidal Love number (Love), and quadrupole moment ($Q$). The universal relations among them (I-Love-Q relations)
can be used to reduce the number of parameters in the gravitational-wave templates.
We here study I-Love-Q relations for more realistic white dwarf models than used in previous studies. 
In particular, we extend previous works by including (i) differential rotation and (ii) 
internal temperature profiles taken from detailed stellar evolution calculations.
We use the publicly available stellar evolution code MESA to generate cooling models of both low- and high-mass white dwarfs. 
We  show that differential rotation causes the I-Q relation (and similarly the Love-Q relation)  to deviate from that of constant rotation. 
We also find that the introduction of finite temperatures causes the white dwarf to move along the zero-temperature mass sequence of I-Q values, moving towards values that suggest a lower mass. 
We further find that after only a few Myrs, high-mass white dwarfs are well-described by the zero-temperature model, suggesting that the relations with zero-temperature may be good enough in most practical cases. Low-mass, He-core white dwarfs with thick hydrogen envelopes may undergo long periods of H burning which sustain the stellar temperature and allow deviations from the I-Love-Q relations for longer times.

\end{abstract}




\def\lesssim{\mathrel{\hbox{\rlap{\hbox{\lower5pt\hbox{$\sim$}}}\hbox{$<$}}}}
\def\gtrsim{\mathrel{\hbox{\rlap{\hbox{\lower5pt\hbox{$\sim$}}}\hbox{$>$}}}}

\section{Introduction}

The Laser Interferometer Space Antenna (LISA) mission is slated to launch in the 2030s, and it is expected that LISA will detect thousands of short-period white dwarf-white dwarf (WD-WD) binaries in the galaxy via their gravitational wave (GW) emission \citep{many_wds_lisa}. These binaries will be detected by fitting model waveforms to the observed GW data to infer relevant parameters such as orbital periods, chirp masses, and distances. The GW signal of a WD-WD binary is given by a point-mass contribution and small corrections due to the finite size of the WDs. These small corrections may be measurable for binaries with sufficiently small separations \citep{shah_and_nelemans1, shah_and_nelemans2}.

The leading-order finite-size correction to the GW signal comes from the transfer of angular momentum from the orbit to the spins of the individual WDs by tidal friction. In the limit of strong tidal torques, the spins of the individual WDs may be nearly synchronized to the spin frequency of the orbit well before merger \citep{tidal_heating, new_piro}. The strength of this correction may be estimated by the small parameter $(I_1+I_2)/\mu a^2$ for perfect synchronization, where $I_{1,2}$ is the moment of inertia of each WD, $\mu$ is the reduced mass of the binary, and $a$ is the semimajor axis of the binary orbit \citep{benacquista}. Thus, the moment of inertia of both WDs enter as parameters into the GW signal from a binary.

Higher-order corrections to the GW signal appear due to the quadrupole moments of the individual stars themselves. As each individual star is distorted by tides and rotation, the orbital potential energy is changed, and this alters the relationship between semimajor axis and frequency away from the usual Keplerian one \citep{Poisson:1997ha,ns_conservative_effect, benacquista}; this is often referred to as the \emph{conservative} effect. Additionally, as seen from a non-rotating frame, the quadrupoles raised on each star by tides vary over the orbit, leading to GW radiation emitted from the WDs themselves; this is often referred to as the \emph{non-conservative} or \emph{dissipative} effect \citep{ns_conservative_effect}. Though the contributions from these effects to the GW signal are small, they are likely to be measurable over the lifetime of LISA for systems with high signal-to-noise ratio, a low-mass (large-radius) primary, and a high-mass companion. Thus, we have that the moment of inertia ($I$), tidal Love number (Love), and rotational quadrupole moment ($Q$) of each WD enter as important parameters into the gravitational waveform for a double WD binary.

In the case of neutron stars (NSs), \cite{ilq_og,second_ILQ_paper} found that $I$, Love, and $Q$ were related to each other in a way that was independent of the assumed equation of state (EoS), which is currently poorly understood for NSs. These relationships formed the so-called ``I-Love-Q relations", and they serve to break the degeneracy between $Q$ and other NS parameters, which in turn reduces the overall measurement uncertainties on the latter. Indeed, such universal relations, together with similar relations \citep{binary_love_relations, approx_universal_relations}, have been applied to GW170817 by the LIGO/Virgo Collaborations \citep{measuring_NS_MR_using_ILQ, GW170817}, which helped to improve our understanding of nuclear physics by constraining the relation between pressure and density at supra-nuclear densities.

While different EoS prescriptions in the NS case may lead to significant differences in the mass-radius relation, the WD EoS is better understood. The EoS of the WD core is a degenerate electron gas, allowing for arbitrarily relativistic electron velocity, and there are small corrections due to the electrostatic interaction of the electrons and ions, as well as due to finite temperature. Outside the degenerate core, there is a non-degenerate envelope composed of hydrogen and/or helium. The size of the envelope may be larger for low-mass WDs, and for masses less than $\la 0.2\, M_\odot$, residual nuclear burning in a thick hydrogen envelope may significantly slow the WD's cooling. The composition of the WD is determined by post-main sequence nuclear burning as well as binary mass transfer episodes.   Low-mass WDs ($M \la 0.4\, M_\odot$) are comprised of a He core and H-rich envelope, while the bulk of the higher mass WD will have a mixture of Carbon and Oxygen in the core, with thin shells of He and H outside. Element diffusion acts to allow heavy elements to sink down and light elements to rise up during the evolution, leading to the fractionated stucture. The starting central temperature on the WD cooling track is determined by post-main sequence He core burning and H/He shell burning.  
The finite initial core temperature, relative to the central Fermi energy, allows some thermal pressure support, causing the WD to deviate from the zero-temperature solution. Thus, in the case of WDs, the I-Love-Q relations aim to relate relevant parameters of WDs to each other across varying internal structure (such as varying composition, rotation profile, central temperature, age, etc.) as was accomplished in the NS case.

In the case of binary WDs, it is $I$, Love, and $Q$ that encode the effects of the finite sizes of the WDs onto the gravitational waveform (unlike the binary NS case where it is the spin angular momentum instead of $I$ that is measurable). \cite{ilq_wd} first studied these I-Love-Q relations in the context of WDs and found that differences in WD compositions, and hence mean mass per electron and Coulomb interaction effects, did not affect the relationships between $I$, Love, and $Q$. \cite{hotwd} further studied WDs with finite and uniform temperature, finding that finite temperature effects did cause the relations to deviate from the zero-temperature result.

In this work, we further investigate these I-Love-Q relations for more realistic WD models. In particular, we consider two main extensions, (i) differential rotation and (ii) self-consistent temperature profiles.

As WDs in close binaries may be subject to strong tidal torques and internal angular momentum redistribution mechanisms, their rotational angular frequency profiles may be far from uniform. We study how this may affect the I-Love-Q relations for WDs. \cite{why_ILQ} have already studied these relations in the context of differentially-rotating main sequence stars (not compact objects), and we build upon this work by studying how differential rotation affects WD I-Love-Q relations\footnote{See also \citealt{Bretz:2015rna} for the universal relations among multipole moments of differentially-rotating Newtonian polytropes.}. To do so, we solve for the modification to interior structure using the Hartle-Thorne formalism \citep{realhartle,1968ApJ...153..807H} in which we treat the stellar rotation as a small perturbation. The results are compared to the uniform rotation case (see \citealt{ilq_wd}). True differential rotation that occurs in the interiors of WDs is a complicated process;
in this work, we assume a parametrized model and investigate possible deviations from the usual I-Love-Q relations that could occur.

Additionally, we will be studying the effects of finite temperature on WD I-Love-Q relations, extending the previous work \citep{hotwd} by using the publicly available MESA code \citep{mesa} that evolves a star from the pre-main sequence to the WD cooling track and allows for nuclear burning, as well as convective and radiative heat transport. We generate two WD models, a low-mass He-core WD with mass $M=0.15\, M_\odot$ and a more massive C/O-core WD with mass $M=0.83\, M_\odot$ to test if the I-Love-Q relations hold in these two extremes. The use of MESA models allows two improvements over the uniform temperature used by \cite{hotwd}. First, the core temperature is set by post-main sequence burning and is not a free parameter. Second, while WD cores are nearly isothermal a few thermal times after formation, their envelopes have a steep outward temperature gradient, and hence for low mass WD with thick envelopes the uniform temperature assumption may overestimate thermal support in the envelope. The finite temperature effects are most pronounced in low mass WD with thick H envelopes, which may have residual nuclear burning for Gyrs which delays the cooling of the WD. 

The Newtonian equations of inviscid fluid motion will be used throughout this paper, although occasional contact is made with results for neutron stars, which used General Relativity.

The remainder of the paper is organized as follows.
In Section \ref{sec:diffrot}, we show how differential rotation alters the WD I-Love-Q relations compared to that of constant rotation. In Section \ref{sec:hotwd}, we analyze the effects of finite temperature on the structure of WDs and how the I-Love-Q relations are affected. Finally, Section \ref{sec:conclusion} contains a discussion of the the results and the conclusions.

\section{Differential Rotation} \label{sec:diffrot}

\subsection{The background model} \label{sec:background}

For the study of differential rotation in the present section (\ref{sec:diffrot}), it is convenient to employ simple models of zero temperature WDs.
The background model is constructed by solving the equations of hydrostatic balance, interior mass, and the equation of state. 
The equation of state, or pressure-density relation, which is assumed here is $P=P_e + P_c + P_{\rm TF}$, where $P_e$ is the pressure from degenerate electrons and $P_c$ is due to the electrostatic attraction among electrons and nuclei, and $P_{\rm TF}$ is the correction due to non-uniform electron density in each ion cell. The dominant contribution to the pressure is from the degenerate electron gas
(e.g. \citealt{book}):
\beq
\label{degen_pressure}
P_e = \frac{m_e c^2}{8 \pi^2 \lambda_e^3} \left[ x \Big(1+x^2\Big)^{1/2}\Big(\frac{2}{3}x^2-1\Big) + \ln\Big(x + (1+x^2)^{1/2}\Big)\right],
\eeq
where $m_e$ is the mass of the electron, $c$ is the speed of light, $\lambda_e \equiv \hbar/m_ec$ is the electron Compton wavelength, and $x \equiv p_F/m_ec \ll 1$ for the non-relativistic case and $x \gg 1$ for the relativistic case. The Fermi momentum $p_F$ is related to the mass density $\rho$ by:
\beq
p_F = \left( \frac{3h^3}{8\pi} \frac{\rho}{\mu_e m_p}\right)^{1/3},
\eeq
where $\mu_e$ is the mean mass per electron ($\mu_e=A/Z$ for a gas with one ion of mass $Am_p$ and charge $+Ze$) and $m_p$ is the mass of the proton. For stellar masses well above Jupiter's mass, the Coulomb and Thomas-Fermi corrections are a small perturbation, given by 
(see e.g. \citealt{1961ApJ...134..669S}):

\begin{align}
    P_c + P_{\rm TF} = -\frac{m_e c^2}{\lambda_e^3} \Bigg[ &\frac{\alpha Z^{2/3}}{10 \pi^2} \left(\frac{4}{9\pi}\right)^{1/3}x^4 \nn \\
    &+ \frac{162}{175} \frac{(\alpha Z^{2/3})^2}{9\pi^2} \left(\frac{4}{9\pi}\right)^{2/3} \frac{x^5}{\sqrt{1+x^2}}\Bigg],
\end{align}
where $\alpha = 1/137$ is the fine-structure constant.
We see that the Coulomb correction is proportional to $\rho^{4/3}$, while the degeneracy pressure is proportional to  $\rho^{5/3}$, indicating that the Coulomb and Thomas-Fermi corrections become smaller as WD central density increases.

In this section, the WD is assumed to be made of a single ion, and models with either $^4$He, $^{12}$C or $^{16}$O will be given. Some results are presented including $P_C+P_{\rm TF}$ while others ignore these corrections. 
The purpose of this section is not to perform an in-depth analysis of WD composition. 
Rather, the goal is to explore the effect of differential rotation on WD models with different mass and composition. Similarly, we choose to calculate WD parameters outside of their physical mass ranges (C/O WDs do not exist below roughly 0.45 $M_{\odot}$) 
to show that the I-Love-Q relations hold for WDs even in these unphysical regimes.


\subsection{Differential rotation profile} \label{sec:parametrized}
The equation of hydrostatic balance is given by
\beq
\label{hse_original}
0 = -\vgrad P - \rho \vgrad \Phi - \rho \vgrad U,
\eeq
where $P$ and $\rho$ are the pressure and density,
$\Phi$ is the gravitational potential
and $U$ is the perturbing potential, here due to the centrifugal force.
For constant rotation, the perturbing potential is given by
\beq
U = -\frac{1}{2} \Omega^2 \varpi^2,
\eeq
where $\Omega$ is the spin frequency of the star, $\varpi = r\sin{\theta}$ is the cylindrical radius,  $\theta$ is the colatitude and $r$ is the spherical radius. The resulting centrifugal force is
\beq
\bm{F_c} = -\vgrad U = \Omega^2 \varpi \bm{\Hat{\varpi}}.
\eeq

We assume the following form for $\Omega=\Omega(\varpi)$ (see \citealt{form_of_omega}):
\beq
\label{def_omega}
\frac{\Omega}{\Omega_c} = \frac{A^2}{A^2 + \varpi^2}.
\eeq
Here, $\Omega_c$ is the central rotation frequency and $A$ may be thought of as a core radius of the rotation profile. The $A\rightarrow\infty$ limit recovers constant $\Omega$, while small but nonzero $A$ gives constant specific angular momentum $j=\Omega \varpi^2$. 
Demanding that the centrifugal force still be of the same form
\beq
F_c = -\frac{dU}{d\varpi} = \Omega^2 \varpi,
\eeq
gives the following potential for the differentially rotating case
\begin{align}
    U &= -\int_0^\varpi \Omega_c^2 \left( \frac{A^2}{A^2 + \varpi^2} \right)^2 \varpi d\varpi \\
    \label{U}
    &= -\frac{1}{2} \Omega_c^2 \frac{\varpi^2}{1 + \varpi^2/A^2}.
\end{align}
Written in this form, it is clear that in the limit that $A \gg \varpi$, we recover the results of constant rotation. In Appendix \ref{sec:derive_eqns}, 
the perturbed structure equations are given including the effect of differential rotation.

Since the chief purpose of this work is to aid in modelling gravitational waveforms from WD binaries, we only consider leading-order contributions to $I$, Love, and $Q$ in rotation. These terms appear in small corrections to the waveform, so spin corrections to these terms  are higher-order and therefore negligible. Due to this assumption, the moment of inertia is entirely a background quantity, unaffected by (differential) rotation while the quadrupole moment is proportional to spin squared. The tidal Love number is also unaffected by rotation, which we explain in more detail in Section \ref{sec:def_lovenum}.




\subsection{Choice of $A$ and $\Omega_c$} \label{sec:fixJ}

The core radius for differential rotation is expressed as a dimensionless parameter $A_s\equiv A/R$, where $R$ is the radius of the non-rotating background star. Only a certain range of $A_s$ is physically relevant; if $A_s \ll 1$, then nearly the entire star has constant-$j$ rotation, and the entire star has uniform rotation for $A_s \gg 1$, a case already studied (see \citealt{ilq_wd}).
Thus, any study of the effects of differential rotation on the I-Love-Q relations need only concern itself with intermediate values of $A_s$. 
Models are presented over a range of $A_s$ between 0.1 and 10
as well as for a range of WD central densities. We then calculated $I$, $Q$, and the tidal Love number for each model using 
the perturbative approach given in Appendix \ref{sec:other_vars}.

As we now motivate, sequences of models with fixed $J$ will be used in order to study the variation of $I$ and $Q$ for different $A_s$. 
In previous works, it was natural to fix the spin frequency of the star at the breakup frequency $\sqrt{GM/R^3}$ 
as this demonstrated the maximum possible effect of rotation. Here, however, the free parameter is the central spin frequency $\Omega_c$, which is different from the spin frequency at the surface. 
We fix each WD's value of $J$ to a specified $J_\mathrm{fixed}$ by adjusting the value of $\Omega_c$.
Each WD structure was first computed using a test value of $\Omega_c$ equal to $\sqrt{G M/R^3}$, and its angular momentum $J_0$ was calculated using the methods of Section \ref{sec:other_vars}. The value of $\Omega_c$ was then scaled down by a factor of $J_\mathrm{fixed}/J_0$, since $J \propto \Omega_c$ at leading order in spin. The model was then computed again using this new value of $\Omega_c$, and the parameters of this second iteration were recorded. Had we not chosen to fix $J$ and instead fixed $\Omega_c$, the sequences of $I$ and $Q$ would not be physically meaningful, e.g. for small $A_s$, most of the star would be rotating slowly.
By fixing $J$, we aim to compare similar stars to each other, rather than stars rotating at significantly different rates, and by doing so properly calibrate the effects of differential rotation. 

\subsection{The Love number} \label{sec:def_lovenum}

In Newtonian physics (but not in full General Relativity -- see \citealt{love_nums}), the Love number represents the linear response of the star to a perturbing potential, and is simply related to the quadrupole moment. In the case of tides, 
the point mass gravity of star 2 gives rise to a quadrupole tidal potential acting on star 1
\beq
U_2(\vec{x}_1) = -G m_2 \frac{r_1^2}{r_2^{3}} P_2(\cos\theta_{12}).
\eeq
Here $\vec{x}_1=(r_1,\theta_1,\phi_1)$ are the coordinates inside star 1. The coordinates $(r_2,\theta_2,\phi_2)$ describe the position of the center of mass of star 2 as seen from the center of mass of star 1. The angle $\theta_{12}$ is defined by $\cos(\theta_{12})=\vec{x}_1 \cdot \vec{x}_2/(r_1r_2)$. This tidal potential will cause density changes within star 1, causing the external potential to deviate from the point mass value through the quadrupole moment
\beq
Q_1 = \lambda_1 m_2 \frac{R_1^5}{r_2^3},
\eeq
where $\lambda_1$ is the quadrupolar Love number of star 1, a dimensionless number mainly dependent on the central concentration.
The case of uniform rotation is similar. The quadrupolar centrifugal potential is
\beq
U(\vec{x}_1) = \frac{1}{3}  \Omega_1^2 r_1^2  P_2(\cos\theta_1).
\eeq
This potential has the same form as for tides, and so the Love number, which is independent of any constants in $U$, must be the same
as for tides. The quadrupole moment is then only different due to the parameters in the forcing potential, and an extra factor of $P_2(0)=-1/2$ for rotation axis perpendicular to the orbital plane, giving
\beq
Q_1 = \frac{1}{6G}  \lambda_1 \Omega_1^2 R_1^5.
\label{eq:rotation_quadrupole}
\eeq
Hence in the Newtonian case, there is a simple relationship between Love numbers and quadrupole moments, at least for uniform rotation, and it is not necessary to consider the full I-Love-Q relations.

Different branches of physics and astronomy refer to different quantities by the term ``Love number". Here
the Love number is defined by a ratio of response potential to forcing potential  for a particular spherical harmonic component and evaluated at the surface,
\beq
\lambda  \equiv  \frac{\delta \Phi_\ell(R) }{U_\ell(R)}
\eeq
(also see Appendix \ref{sec:other_vars}). In this work, the main perturbing potential we are dealing with is the centrifugal potential. However, in the chief application of this work (GWs from binary WDs), it is rather the tidal potential (and thus the tidal Love number) which is more important (see \citealt{benacquista}). This is the ``Love" of the I-Love-Q relations, as it provides information about small corrections to the background point-mass GW signal emitted by a binary system.

In Newtonian physics assuming constant rotation, the rotational Love number $\lambda_R$ and the tidal Love number $\lambda_T$ are equivalent \citep{love_nums}, and they are often used interchangeably. The tidal Love number depends on the type of material being tidally distorted; thus, WDs (which vary in polytropic index from 3/2 to 3) vary in tidal Love number depending on their density profile. However, to leading order in spin, the tidal Love number does not depend on the amount of rotation occurring inside the WD, so there should be no dependence of the tidal Love number on the amount of differential rotation occurring inside the WD.

On the other hand, the rotational Love number does depend on the amount of rotation. Let us estimate how $\lambda_R$ scales with $A$.
We begin by looking at the response potential (see Equation \ref{defQ}):
\beq
\Phi^{(2)}_2 \Big|_R \propto Q.
\eeq
In addition to calculating $Q$ using the perturbed potential as in Equation \ref{defQ}, one can integrate to find $Q$ directly:
\beq
Q \propto \int_0^R dr r^4 \delta \rho,
\eeq
where $\delta \rho$ is the $\ell = 2$ mode of the perturbed density profile due to rotation (see Equation \ref{defQ_fluid}) using Lagrangian perturbation theory (Appendix \ref{sec:l0_l2_eqns}). 
From Equations \ref{solve_xir_l2} and \ref{delta_rho}, we know that
\beq
\delta \rho = -\frac{d\rho}{dr} \xi_r = \frac{d\rho}{dr} \frac{U_2}{g},
\eeq
where $U_2$ is the $\ell = 2$ 
forcing potential given in Equation \ref{U2}. Combining all the terms depending on $r$,
the rotational Love number is roughly given by
\begin{align}
    \label{love_estimate}
    \lambda_R = \frac{\Phi^{(2)}_2}{U_2} \Bigg |_R = \frac{\int_0^R h(r) f_2(\alpha) dr}{f_2(\alpha_R)},
\end{align}
where $h(r)$ is a function of $r$, $\alpha \equiv r/A$, $\alpha_R \equiv R/A$, and the function $f_2(\alpha)$ is defined in Equation \ref{f2}. Note that $\lambda_R$ is independent of $\Omega_c$, and it only depends on rotation via the core radius $A$.

We now take two limits of $f_2(\alpha)$.
In the limit of small $A_s$ (large $\alpha$), the function $f_2(\alpha)$ is approximately
\beq
\label{ln_scaling}
f_2(\alpha \gg 1) \sim \frac{1}{\alpha^4} [3 - 2\ln (2\alpha)].
\eeq
Due to the logarithm, we cannot 
pull all $A$ terms out from the integral in Equation \ref{love_estimate}.
%
In the limit of large $A_s$ (small $\alpha$; approaching the constant rotation limit),
\beq
\label{f2_lowalpha}
f_2(\alpha \ll 1) = -\frac{2}{3} + \frac{16}{21} \alpha^2 + \mathcal{O}(\alpha^4),
\eeq
so the correction to the Love number away from its constant-rotation value vanishes on the order of $1/A_s^{2}$.
\begin{figure}
    \centering
    \includegraphics[width=\columnwidth]{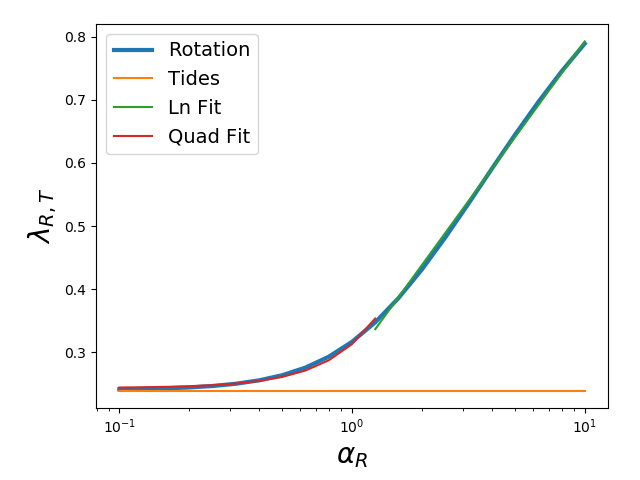}
    \caption{The rotational Love number as a function of the dimensionless radius $\alpha_R \equiv R/A$ at fixed $M =0.6 M_{\odot}$ (blue). Observe how the rotational Love number approaches the tidal Love number (orange) as we decrease $\alpha_R$. The green line labeled ``ln fit" and the red line labeled ``quad" give the large and small $\alpha_R$ approximations, respectively.}
    \label{fig:lambda_vs_alpha_wfit}
\end{figure}

We affirm our above analytical estimates in Figure \ref{fig:lambda_vs_alpha_wfit}. We show the rotational Love number at fixed mass and varying $\alpha_R \equiv R/A$. At small $\alpha_R$, the rotational Love number approaches the tidal Love number, as we expect from Equation \ref{f2_lowalpha}. At large $\alpha_R$, Equation \ref{ln_scaling} does not give an exact scaling with $A$, but we estimate some logarithmic dependence. We attempted a fit to the data in the limit of $\alpha_R > 1$ using the model $\lambda_R = a + b\ln{\alpha_R}$, shown in green. We find good agreement between our fit and the data, confirming that $\lambda_R$ goes logarithmically in the limit of large $\alpha_R$. The parameters $a$ and $b$ depend on the fixed mass, e.g.  $a = 0.29$ and $b = 0.22$ for $M = 0.6 M_{\odot}$. We performed a similar fit to the low-$\alpha_R$ data, using the model $\lambda_R = \lambda_T + c\alpha_R^2$, and we find that this models the data well for $\alpha_R < 1$. As in the high-$\alpha_R$ limit, the parameter $c$ varies according to the fixed mass; for a $0.6 M_{\odot}$ WD we find that $c = 0.070$.

\begin{figure}
    \centering
    \includegraphics[width=\columnwidth]{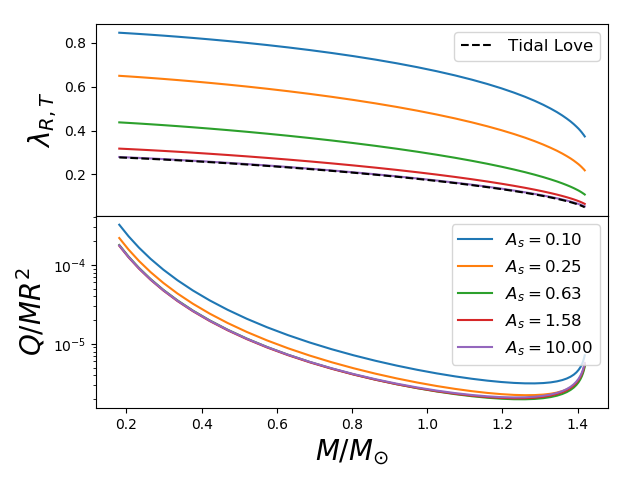}
    \caption{The rotational Love number of differentially-rotating WDs (top) and the dimensionless quadrupole moment (bottom) as functions of WD mass, with the different lines representing different sizes $A_s$ of the uniformly rotating region. Along with the rotational Love numbers, we show the tidal Love number (which at leading order is independent of rotation) for reference. Coulomb corrections are ignored for simplicity.
    As $A_s$ becomes larger than one, we return to the limit of constant rotation, where the rotational Love number and tidal Love number are equal. We see that over a span of WD masses, the normalized quadrupole moment is modestly affected by the varying core radius.}
    \label{fig:QLove_vs_M}
\end{figure}

We now show in the top panel of Figure \ref{fig:QLove_vs_M} rotational Love numbers as functions of WD mass across several values of $A_s$. We also include the tidal Love number as a function of mass as well, for reference. We find that the rotational Love number varies greatly across different $A_s$.
For the chief application of this work (WD GW analysis), we wish to find relations involving the tidal Love number, rather than the rotational Love number. 
Throughout the remainder of this paper, any mention of the Love number refers to the tidal Love number, which is independent of the amount of differential rotation at leading order.

\subsection{The rotationally-induced quadrupole moment}
We next study how the rotationally-induced quadrupole moment $Q$ depends on differential rotation. To illustrate this, we present in the bottom panel of Figure \ref{fig:QLove_vs_M} how the quadrupole moment varies along the sequence of possible WD masses with varying $A_s$. Because we are only keeping to leading-order in spin, $I$ and $\lambda_T$ are entirely background terms, so it is only $Q$ which varies due to differential rotation. Across the physically reasonable range of $A_s$, there is some modest deviation of $Q/MR^2$ away from the constant-rotation sequence (shown by the overlapping lines for larger values of $A_s$). 

As we discussed in Section \ref{sec:fixJ}, the angular momentum of each system (across varying $A$ and $M$) was fixed to the same value. As we will show later, the value of $J_\mathrm{fixed}$ is irrelevant to the I-Love-Q relations; however, it is of some relevance in Figures \ref{fig:QLove_vs_M} and \ref{fig:scaledQLove_vs_scaledI}, so we shall describe here how we fixed $J$. We wished to avoid creating a WD which rotates at its center faster than its global breakup frequency (avoiding $\Omega_c > \sqrt{GM/R^3}$). Thus, we chose for our value of $J_\mathrm{fixed}$ the value which corresponds to (i) the lowest-mass WD we considered, (ii) with the lowest value of $A_s$ we considered, (iii) rotating centrally at its breakup frequency ($\Omega_c = \sqrt{GM/R^3}$). Here, the lowest mass and $A_s$ we considered were 0.18 $M_{\odot}$ and 0.1.

\begin{figure}
    \centering
    \includegraphics[width=\columnwidth]{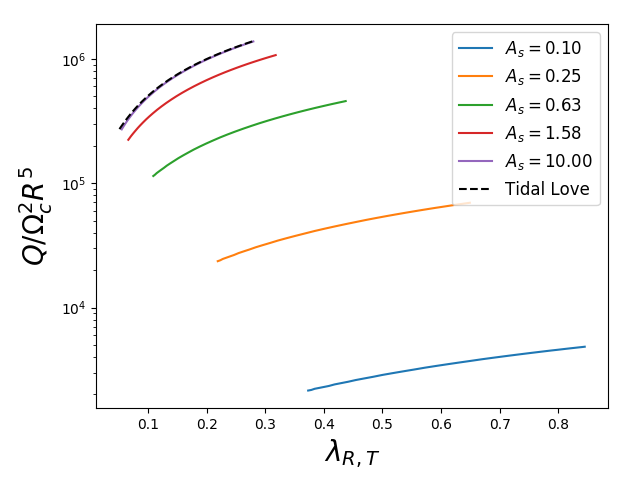}
    \caption{ 
    Rotational Love number $\lambda_R$ vs the appropriately scaled quadrupole moment. For uniform rotation, $Q$ and $\lambda_R$ are related by a constant factor (see Equation 
    \ref{eq:rotation_quadrupole}); however, now that we have promoted the perturbing potential $U$ to be a function of $A$, there is an additional parameter necessary to relate $Q$ and $\lambda_R$ (see Equations \ref{defk2} and \ref{defQ}). For reference, we also present the relation between the rescaled tidally-induced $Q$ and the tidal Love number $\lambda_T$. Observe how the relations for $\lambda_R$ approach that for $\lambda_T$ as one increases $A_s$ (as one approaches the constant rotation case).}
    \label{fig:just_QloveR}
\end{figure}

How are Love numbers and quadrupole moments related?
For uniform rotation the relation is given by Equation \ref{eq:rotation_quadrupole}.
However, in the case of differential rotation, the forcing potential $U$ and hence the rotational Love number will depend on $A$ 
(see Equation \ref{U} and Appendix \ref{sec:def_lovenum}).
Thus, across several values of $A_s$, $\lambda_R$ and $Q$ are instead no longer related by a constant factor. We show this relation in Figure \ref{fig:just_QloveR}. Because the chief application of this work is to GWs where $\lambda_R$ does not play a primary role, we do not discuss this further.

\subsection{I-Love-Q under differential rotation}
\begin{figure}
    \centering
    \includegraphics[width=\columnwidth]{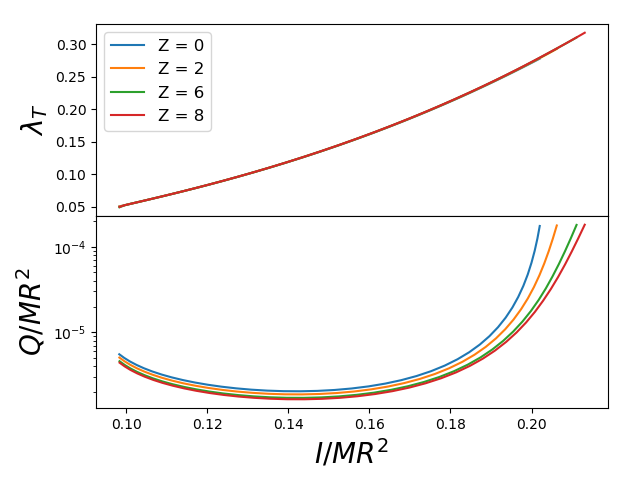}
    \caption{
    Normalized $I$ vs the tidal Love number (top) and normalized $Q$ (bottom) across a range of WD masses, now at constant $A_s = 1$ and across several different WD compositions. We find here that the relationship between $I$ and $\lambda_T$ is independent of composition. We also find that the relationship between $I$ and $Q$ depends slightly on the composition of the WD chosen across all WD masses. We have fixed $J = J_0$ across the entire range of WD masses, where $J_0$ is given by the angular momentum of the lowest-mass WD with $A/R = 10^{-3}$ and $\Omega_c = \Omega_b$, where $\Omega_b$ is the breakup frequency of the WD.} 
    \label{fig:scaledQLove_vs_scaledI}
\end{figure}

Having all the ingredients at hand, we now study the I-Love-Q relations.
Previous works have shown the composition-independence of the I-Love-Q relations \citep{ilq_wd}, where it was demonstrated that WDs models with and without Coulomb corrections in the EoS
followed the same set of I-Love-Q relations. We wish to show a similar set of relations for differentially-rotating WDs at fixed $J$ across both varying $A$ and composition. As we have stated before, we only keep to leading order in spin, so only $Q$ will vary due to varying $A$. We now fix the scale length $A$ but vary the mass and composition
in Figure \ref{fig:scaledQLove_vs_scaledI}, scaling $Q$ and $I$ so to make them dimensionless. We find that, at fixed $A_s$, the dimensionless $I$ is related to $\lambda_T$ independently of composition, but the same is not true of dimensionless $I$ and dimensionless $Q$.

\begin{figure}
    \centering
    \includegraphics[width=\columnwidth]{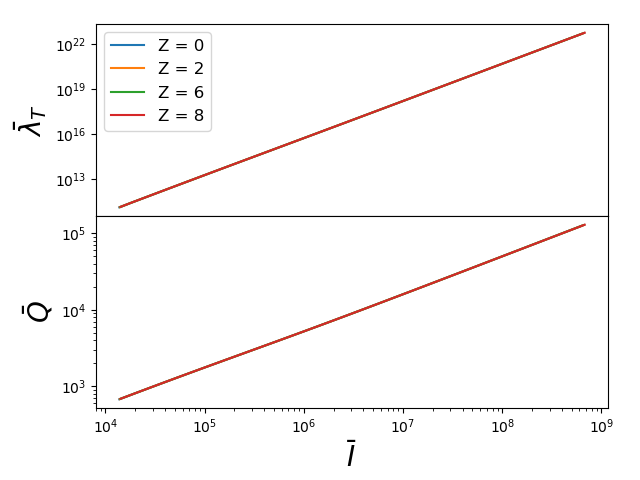}
    \caption{Similar to Figure \ref{fig:scaledQLove_vs_scaledI} but for an alternative choice of dimensionless quantities $\bar I$, $\bar \lambda$ and $\bar Q$ (Equations \ref{Ibar}, \ref{lovebar}, and \ref{Qbar}). Observe that the relations are now universal at fixed $A_s=1$ among various WD compositions in both panels, with less than 1 percent deviation in the I-Love relation and less that 0.4 percent deviation in the I-Q relation.}
    \label{fig:LovebarQbar_vs_Ibar_manyZs}
\end{figure}

Previous works also found that simply normalizing $I$ and $Q$ by $MR^2$ did not make the relations composition-independent (see \citealt{ilq_wd}). We define
\begin{align}
    \label{Ibar}
    \bar{I} &\equiv \left(\frac{c^2}{G}\right)^2 \frac{I}{M^3}, \\
    \label{lovebar}
    \bar{\lambda}_T &\equiv \frac{1}{3} \left(\frac{c^2}{GM/R}\right)^5 \lambda_T,\\
    \label{Qbar}
    \bar{Q} &\equiv \frac{c^2 Q}{J^2/M},
\end{align}
where, $M$, $R$, and $I$ refer to the non-rotating, background component of the star's mass, radius, and moment of inertia (see \citealt{ilq_og}). Similarly, $J$ is only kept to first-order in $\Omega_c$ (see Appendix \ref{sec:other_vars}). Each term is now dimensionless, and importantly, $Q$ has had its $\Omega_c$ dependence scaled out of it (since $Q \sim \Omega_c^2$ and $J \sim \Omega_c$). Previous works have found that comparing these variables to each other results in composition-independent I-Love-Q relations. We show in Figure \ref{fig:LovebarQbar_vs_Ibar_manyZs} these particular scalings of $I$, $\lambda_T$, and $Q$ for differentially-rotating WDs at fixed $A_s$ and angular momentum across a range of WD masses and several different WD compositions. Clearly, these particular scalings result in composition-independent relations at a fixed value of $A_s$.

\begin{figure}
    \centering
    \includegraphics[width=\columnwidth]{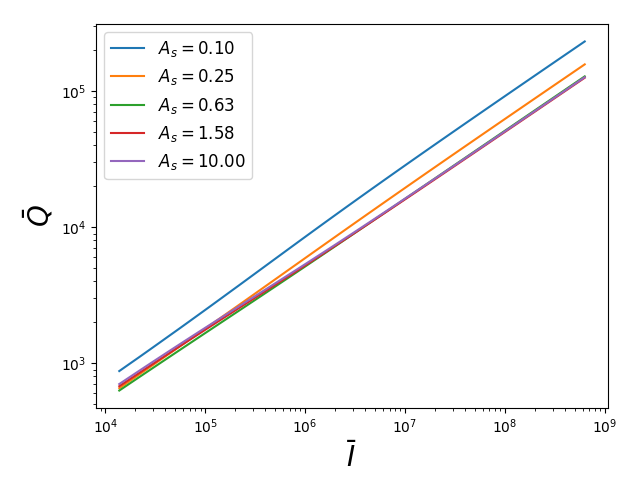}
    \caption{
    The dimensionless $\bar{I}$ vs $\bar{Q}$ across a range of WD masses, using the Chandrasekhar EoS, varied across several core radii. Across all physically reasonable values of the core radius (between 0.1 and 10 times the background radius of the WD), the I-Q relation is no longer universal due to significant dependence on the core radius.}
    \label{fig:LovebarQbar_vs_Ibar_manyAs}
\end{figure}

We now wish to know if these same composition-independent relations hold for differing values of the $A_s$. Because we showed in Figure \ref{fig:LovebarQbar_vs_Ibar_manyZs} that the above scalings are composition-independent, it does not matter which composition we choose, so we ignore Coulomb corrections ($Z=0$).
We show in Figure \ref{fig:LovebarQbar_vs_Ibar_manyAs} these same scalings of $I$ and $Q$ across a range of WD masses, but now we choose one composition and select a variety of core radii. $\bar{I}$ and $\bar{\lambda}_T$ are related in a way independent of $A_s$ by assumption, but the same is not true for $\bar{Q}$. Thus, differential rotation has introduced a new degree of freedom to the original I-Love-Q relations for WDs with constant rotation.

\subsection{A new relation} \label{sec:fitting}
\begin{figure}
    \centering
    \includegraphics[width=\columnwidth]{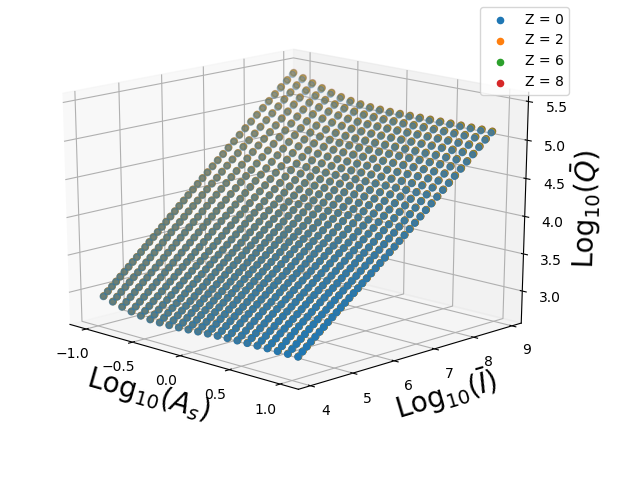}
    \caption{The fundamental plane of the logarithms of $\bar{I}$, $\bar{Q}$, and the scaled core radius $A_s$. The plane is insensitive to the composition of the WD chosen. Our fit to this plane is accurate to within 2\% across the range of core radii we considered. The form of the fit is in Equation \ref{form_of_fit}, and the coefficients are listed in Table \ref{tab:fitting_parameters_cold}.}
    \label{fig:fundamental_plane}
\end{figure}

We have shown in Figure \ref{fig:LovebarQbar_vs_Ibar_manyAs} that the constant-rotation I-Q (and similarly Love-Q) relation does not hold in differentially-rotating WDs to leading order in rotation, acquiring an additional degree of freedom (though it is still universal under variation in compositions). By assumption, the I-Love relation is independent of rotation at leading order.
We therefore seek a new relation, one that accounts for the presence of differential rotation; we seek an $I-Q-A_s$ relation. If there were no dependence on $A_s$, the relation between $\bar{I}$ and $\bar{Q}$ would be a straight line in log space. However, now that there is some $A_s$ dependence, it is likely that there may exist some sort of fundamental plane relating the three variables. Indeed, in Figure \ref{fig:fundamental_plane} we show this exact plane of $\bar{I}$, $\bar{Q}$, and the scaled core radius $A_s$ in log space.

We attempted a fit to the data, which we found to be accurate to within 2\% across the range of $A_s$ we considered. The form of the fit is
\begin{align}
    \label{form_of_fit}
    \log_{10}\bar{Q} = &\left[a_1 + \frac{a_2}{\log_{10}A_s - n} + \frac{a_3}{(\log_{10}A_s-n)^2}\right] \\
    &\times \Big\{1-\tanh\Big[a_4(\log_{10}A_s-a_5)\Big]\Big\} + a_6 + a_7\log_{10}\bar{I}, \nn
\end{align}
and the coefficients are listed in Table \ref{tab:fitting_parameters_cold}.

We note that fixing $J$ does not affect the $\bar{I} - \bar{Q} - A_s$ relation. Since $J \propto \Omega_c$ at leading order, we may say that
\beq
J = \Omega_c \times g(A_s),
\eeq
where $g(A_s)$ is some function of the scaled core radius. To fix $J$ at a given value for any central density and core radius, all that needs to be done is to scale $\Omega_c$ up or down accordingly. Then, since $Q \propto \Omega_c^2$, we may say that
\beq
Q = \Omega_c^2 \times h(A_s),
\eeq
where $h(A_s)$ is some other function of the scaled core radius. Then, when we scale $Q$ by $J^2$ to compute $\bar Q$, the dependence on $\Omega_c$ cancels out, and there is only some dependence on $A_s$ left. This relation between $\bar{I}$, $\bar{Q}$, and $A_s$ is therefore independent of the value of $J$ we have fixed.

Let us close this section by commenting on the universality in the I-Love-Q relations.
In the sense that $I$, Love, and $Q$ share the same relation regardless of the composition of the WD, the I-Love-Q relations hold under differential rotation. However, in the sense that $I$, Love, and $Q$ are always related to each other in the same way, then we say that differential rotation has broken the I-Love-Q relations. Rather, it may be most accurate to say that in the case of differential rotation, the original I-Love-Q relations for WDs with constant rotation are incomplete, and we have here made them more complete and applicable to more realistic WDs.
\begin{table}
    \centering
    \begin{tabular}{c|l}
      Variable & Value \\
      \hline
        $a_1$ & $4.127$\\
        $a_2$ & $4.065\times 10^{1}$\\
        $a_3$ & $9.918\times 10^{1}$\\
        $a_4$ & $1.904$\\
        $a_5$ & $-3.655\times 10^{-1}$ \\
        $a_6$ & $7.544 \times 10^{-1}$\\
        $a_7$ & $4.962 \times 10^{-1}$ \\
       $n$ & $5$
    \end{tabular}
    \caption{Fitting constants for the $I-Q-A_s$ relations. The form of the fit is given in Equation \ref{form_of_fit}.}
    \label{tab:fitting_parameters_cold}
\end{table}

\section{Hot White Dwarfs} \label{sec:hotwd}
We now consider a different aspect of the WD I-Love-Q relations, namely the effect of finite temperature. 
Young WDs have sufficient core temperature that 
thermal pressure support may be important.
Additionally, \cite{tidal_heating} showed that in binary systems, tidal friction may heat WD interiors, reversing their natural progression along the cooling track. For WD binaries that are near merger, this tidal heating may significantly raise the temperature of both components \citep{piro}, so many of the systems LISA detects may have larger-than-expected temperatures. We now investigate these finite-temperature WDs and study whether the I-Love-Q relations hold for such objects.

\subsection{Thermal Pressure Support}
In this section, we now relax our previous assumption that the WDs we study are at zero temperature. The effects of finite temperature are present in the EoS. One may characterize the strength of the effects of finite temperature on WDs by the ratio of the thermal energy (given by $kT$) to the Fermi energy of the WD. The Fermi energy may be calculated from the Fermi momentum $p_F$, which is given in terms of the density
\beq
p_F = \left(3\pi^2 \hbar^3 \frac{\rho}{\mu_e m_p}\right)^{1/3},
\eeq
where $\hbar$ is Planck's reduced constant, $\rho$ is the mass density of the WD (dominated by nucleons), $\mu_e$ is the number of nucleons per electron in the WD (2 for most compositions), and $m_p$ is the mass of the proton. The Fermi energy is $E_{\rm F} \simeq p_{\rm F}^2/2m_e$ for densities below $10^6$ g/cm$^3$ and $E_{\rm F} \simeq p_{\rm F}c$ at larger densities.
For a $M=0.15\, M_\odot$ He-core WD soon after formation,
\beq
\label{eq:kT_EF_low2}
\frac{kT_c}{E_f} \approx 0.03 \left( \frac{\rho_c}{10^5 \text{ g/cm}^3} \right)^{-2/3} \left( \frac{T_c}{10^7 \text{ K}} \right).
\eeq
Higher-mass C/O-core WDs ($M \simeq 1 M_{\odot}$ corresponds to roughly $\rho_c\sim 10^8$ g/cm$^3$) may have central temperatures as large as $10^8$ K which raises the ratio by a factor of 2 or so, but such large temperatures are a short-lived state, and the WD will cool down from this value rapidly. Therefore, we expect that it is instead low-mass WDs that will be affected the most by the introduction of finite temperature, and even then only slightly.

Including thermal pressure leads to larger mass and radius, and hence momentum of inertia, at fixed central density.
See \cite{hotwd} and references therein for further discussion of how the introduction of finite temperature changes the parameters of WDs.

We here relax the isothermality assumption imposed in the previous work on the I-Love-Q relations for hot WDs \citep{hotwd} and investigate the regime of WDs in which this assumption is least likely to hold. It has been shown (see e.g. \citealt{book}, Chapter 4) that WDs have an isothermal core, covered by a thin shell of non-isothermal, non-degenerate gas. The fraction of the WD's radius covered by this shell increases as the mass of the WD decreases. Thus, we consider a 0.15 $M_{\odot}$ He WD as our test case for these I-Love-Q relations with finite temperature and a non-isothermal temperature profile.


\subsection{ Perturbations to the Structure at Fixed Mass }
The WD models generated by the MESA code have a certain mass. The Hartle-Thorne formalism to compute rotational perturbations fixes central density and computes mass as a function of rotation rate.
We also include spin corrections to 
order $\Omega^2$ in the global variables $R$, $M$, $I$, and $J$.
For these reasons, it is more convenient to use the Lagrangian perturbation theory (see \citealt{book}, Chapter 6) to solve the equations of interior structure. Under this formalism, the mass of the WD is held constant before and after the perturbative rotation is ``turned on", and the central density is allowed to vary. See Appendix \ref{sec:l0_l2_eqns} for a thorough discussion of this formalism.



\subsection{Details of the He-core MESA Models}
\begin{figure}
    \centering
    \includegraphics[width=\columnwidth]{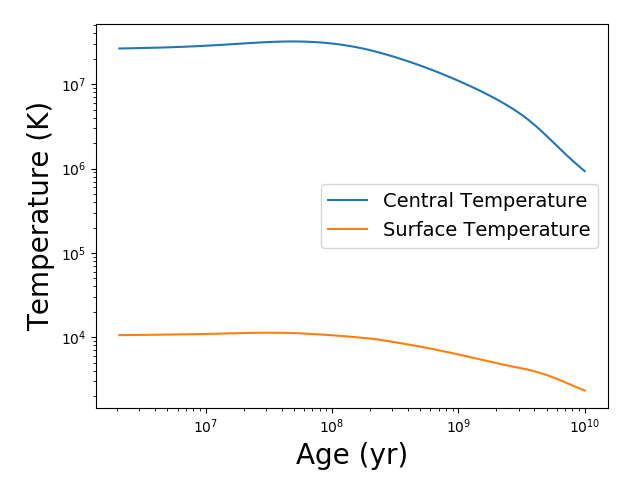}
    \caption{The surface and interior temperature of the cooling WD model for a 0.15 $M_{\odot}$ He WD, generated by the MESA code. In this plot, we have removed all data from before the system reconfigured itself into thermal equilibrium. The WD begins with a central temperature of a few $10^7$ K, and after 10 Gyr, its central temperature has dropped to $10^6$ K.} 
    \label{fig:age_vs_temp}
\end{figure}

To solve the equations of structure
we use the publicly available, stellar-interiors code MESA \citep{mesa}. 
For the He WD model, we modified the ``make\_he\_wd" test\_suite in the MESA package. In this code, MESA begins by evolving a 1.5 $M_{\odot}$ pre-main sequence model until the mass of the interior helium core has reached the specified mass (here 0.15 $M_{\odot}$). The mass of the helium core is defined by the outermost location where the abundance of hydrogen is less than one percent. Next, MESA removes the excess mass ($\Delta M$ = 1.35 $M_{\odot}$) from the outside of the star rapidly, leaving only the helium core. 
The MESA code then ``relaxes" the helium abundance to 99\% over the star and makes all element abundances uniform over the star. Finally, having an appropriately massive He WD, the code allows the WD to cool in isolation. This cooling is slowed, however, by the small amount of hydrogen burning taking place in the center; it is likely that real WDs have larger hydrogen envelopes than the MESA code constructs, which would lead to greater heat generation due to hydrogen burning, ultimately leading to a longer cooling timescale.
We show in Figure \ref{fig:age_vs_temp} the central and surface temperatures of this WD as a function of its age. The WD begins with a central temperature on the order of $10^7$ K, and it decreases to $10^6$ K over 10 Gyr.

\subsection{Results for the He-core MESA Models}

\begin{figure}
    \centering
    \includegraphics[width=\columnwidth, height=3in]{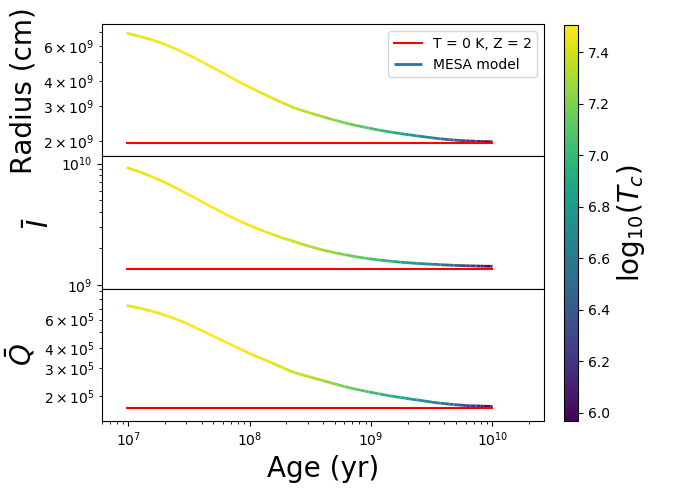}
    \caption{Radius, $\bar{I}$ and $\bar{Q}$ as functions of age for zero-temperature models and the MESA model for a 0.15 $M_{\odot}$ He WD. The data are color-coded to match the central temperature of the WD. The colorbar shows the log of the central temperature in Kelvins. Toward the end of the evolution, the MESA model tends to approach the zero-temperature models (red), as expected.}
    \label{fig:RIQ_vs_age}
\end{figure}

We present here the results of our calculations for hot He WDs. The MESA-generated WD evolved for $10^{10}$ years, and at each time-step, the perturbations to its interior structure due to rotation were calculated (see Appendix \ref{sec:l0_l2_eqns}). We wish to verify the results we obtained from our numerical calculations by comparing the late-time data with the semi-analytic zero temperature WD models which include only degeneracy pressure. In Figure \ref{fig:RIQ_vs_age}, we present the WD's radius, as well as the particular scalings $\bar{I}$ and $\bar{Q}$ (see Equations \ref{Ibar} and \ref{Qbar}) as functions of age. We show data from the MESA calculation, as well as the zero-temperature data for a He WD of the same mass for reference. Over Gyr timescales, the WD cools sufficiently and is well-approximated by the zero-temperature model. The data from the MESA system are color-coded according to the central temperature; the colorbar on the right of the figure shows the log of the central temperature in Kelvin.

Though the numerical MESA data is well-approximated by the zero-temperature result after several Gyrs, deviations do occur at early times. 
Over the first Gyr, the WD cools significantly, and its radius shrinks by a factor of a few. Therefore, we cannot say that the I-Q relation holds in general for hot WDs. However, we estimate that at central temperatures below a few times $10^6$ K, the zero-temperature model will return small enough errors that it is considered a suitable model. In addition, 
the effects of non-zero temperature are 
largest for low-mass WDs (see \citealt{book} and \citealt{hotwd}). The WD we have considered here has a mass of $0.15 M_{\odot}$, among the lowest-mass WDs to have ever been detected \citep{lowmass_wd}. 
For more massive WDs, the central temperature at which the zero-temperature model begins to approximate the system well is higher than we have stated here.

\subsection{Finite-temperature C/O WD} \label{sec:co_wd}
We will now 
investigate 
if more massive WDs are well-modelled by the zero-temperature model at earlier ages and larger temperatures. 
We used the test suite ``make\_co\_wd" in the MESA package, creating a 0.835 $M_{\odot}$ C/O WD that cooled in isolation. Then, as in the He WD case, we solved for the perturbations to the background structure provided by MESA. 

\begin{figure}
    \centering
    \includegraphics[width=\columnwidth]{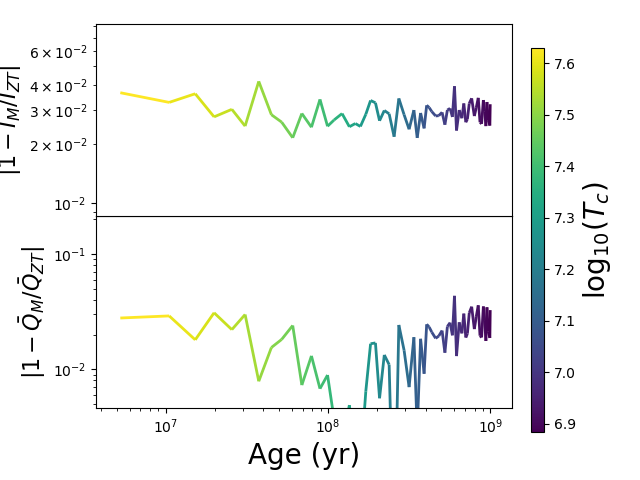}
    \caption{The absolute fractional difference in the quantities $\bar{I}$ and $\bar{Q}$ between the MESA model and the zero-temperature model for a 0.835 $M_{\odot}$ C/O WD as a function of the age of the WD. The meaning of the color-coding is the same as in Figure \ref{fig:RIQ_vs_age}.}
    \label{fig:co_wd_fracdiff}
\end{figure}

We present the results of our calculations in Figure \ref{fig:co_wd_fracdiff}. Unlike Figure \ref{fig:RIQ_vs_age} for a 0.15$M_\odot$ He WD, we show the absolute fractional difference between the MESA model and the zero-temperature model in the quantities $\bar{I}$ and $\bar{Q}$ (see Equations \ref{Ibar} and \ref{Qbar}).
Again, the data are color-coded to match the WD central temperature. The colorbar on the right shows the temperature in units of $\log_{10}$ Kelvin. The WD is born at a very large temperature, since it comes from the core of a roughly $4 M_{\odot}$ star. For higher-mass WDs, the composition 
profile exhibits a carbon/oxygen core, as well a helium/hydrogen envelope. For our zero-temperature models, we assume a constant-composition of Z = 6.

After 10 Myr, the finite-temperature MESA model and the zero-temperature model agree for both $\bar I$ and $\bar Q$ within a numerical error of a few \%.
At this age, the WD has a central temperature of roughly $10^{7.6}$ K, hotter than the He WD core at its genesis. Thus, we have confirmed our hypothesis, that higher-mass WDs are well-modelled by the zero-temperature model at larger temperatures and smaller ages. Since it took this WD tens of Myrs to cool to approximately zero temperature (a short timescale, astronomically speaking), it is likely that most WDs that we have observed will be well-described by the zero-temperature model.

We find that, in the case of high-mass WDs, realistic finite temperatures do not alter the zero-temperature I-Love-Q relations. Up to a few percent error (created by the MESA-evolved WD not being precisely modelled by a constant Z = 6 interior composition), the high-mass WD is well-described by the zero-temperature model in all observable times. 


\subsection{I-Q Relation}
\begin{figure}
    \centering
    \includegraphics[width=\columnwidth]{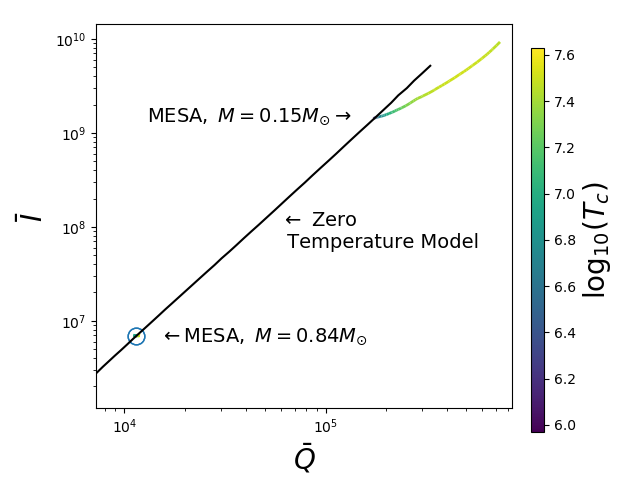}
    \caption{$\bar{I}$ and $\bar{Q}$ over time for both the He WD and the C/O WD we evolved using the MESA code, as well as the same variables over many central densities in the zero-temperature model. Since the composition we choose for our zero-temperature model does not matter \citep{ilq_wd}, we choose the Chandrasekhar EoS for simplicity. The MESA results are shown by lines because the values vary in time due to the WD temperature evolution; the zero-temperature model is a line because the values vary across the many different central densities we show here. Because the C/O WD is not significantly affected by finite temperature (see Figure \ref{fig:co_wd_fracdiff}), its trajectory in $\bar{I}-\bar{Q}$ space is difficult to see. Rather than enlarge it so that it may become visible, we choose to circle the surrounding region to indicate its location while still retaining its point-like appearance. As in Figures \ref{fig:RIQ_vs_age} and \ref{fig:co_wd_fracdiff}, we color-code the MESA models according to their central temperature.} 
    \label{fig:IQbar_MESA_vs_hse}
\end{figure}

In Figure \ref{fig:IQbar_MESA_vs_hse}, we show the time-evolution of $\bar{I}$ and $\bar{Q}$ for both the MESA-evolved He WD and C/O WD, compared to the sequence of $\bar{I}$ and $\bar{Q}$ of zero-temperature WDs across many central densities. We find that the tendency is for the hot WD to begin on the right side of the cold WD $\bar{I}-\bar{Q}$ sequence, then to fall back down to its zero-temperature value along the sequence. This is most noticeable in the He WD; the C/O WD is not noticeably affected by finite temperature, and it appears as a single point in Figure \ref{fig:IQbar_MESA_vs_hse}.

Let us compare the MESA curves and zero-temperature models in more detail. Notice that lower masses correspond to larger $\bar{I}$ and $\bar{Q}$ \citep{ilq_og}. In the case of the higher-mass WD, there is practically no deviation from the cold WD sequence, as we noted previously. However, the low-mass WD deviates significantly from its zero temperature $\bar{I}-\bar{Q}$ point, even after astrophysically long timescales. We notice that the deviation is away from the cold WD sequence and toward generally larger $\bar{Q}$ than $\bar{I}$, as is demonstrated in Figure \ref{fig:RIQ_vs_age}.


\section{Summary and Conclusion} \label{sec:conclusion}
In this work, we have studied the so-called ``I-Love-Q" relations in the context of realistic WDs. The primary application of these relations is to aid in GW analysis by reducing the overall number of parameters necessary to produce model waveform templates. 

We first studied the effect of a parametrized form of differential rotation on these relations, where the amount of differential rotation was characterized by a core radius $A$. We modified the equations of structure governing WD interiors to account for a rotation frequency $\Omega$ that varies with radius. We then solved these equations across a range of WD masses and physical core radii, while holding the angular momentum fixed. We found that the I-Love-Q relations remain universal under variation in compositions even for differentially-rotating WDs. On the other hand, the I-Q and Love-Q relations for such differentially-rotating WDs deviate from those for WDs under constant rotation. Additionally, we found that unlike the constant rotation case, the rotational and tidal Love numbers are not equal under differential rotation. 

Next, we studied how finite temperature affects the WD I-Q relations. We evolved a 0.15 $M_{\odot}$ He WD using the publicly available MESA code \citep{mesa} for $10^{10}$ years as it cooled in isolation. The MESA code tabulated the background data of the WD's interior profile, and we solved for perturbations to this background profile due to constant rotation. 
We found that the WD cools and becomes well-modelled by the zero-temperature approximation over timescales of order Gyrs.
However, at its genesis, the WD was quite poorly-modelled by the zero-temperature approximation, and we say that the I-Q relation does not hold in general for hot WDs.
We then performed a similar treatment to a C/O WD generated by the MESA code. The program generated a 0.835 $M_{\odot}$ WD and cooled it in isolation. We found that the WD was well-described by the zero-temperature model at all astrophysically relevant ages and at central temperatures less than roughly $10^{7.6}$ K, higher than the central temperature of the He WD at its genesis. Thus, we argue that for most intermediate-mass WDs, any deviations from the zero-temperature model are unlikely to be detectable, ignoring any tidal heating that may occur \citep{tidal_heating, piro}, though we note that at sufficiently large temperatures and low masses, the I-Q relation is not the same as its zero-temperature counterpart.

In the case of the He WD, where finite temperature effects were relevant on Gyr timescales, the system deviated in its I-Q relation from the sequence formed by cold WDs of varying masses. Initially, the WD begins with greater $\bar{Q}$ than $\bar{I}$, then as it evolves, it tend to move leftward in I-Q space until it reaches the cold WD sequence. The C/O WD that we considered was unaffected by finite temperature on astrophysically relevant timescales, and there was no noticeable deviation of this system away from the cold WD sequence.

Although the focus of our paper is for WDs, we here comment on how differential rotation, finite temperatures and deviations from chemical equilibrium affect the I-Love-Q relations for NSs. The effect of differential rotation on universal relations\footnote{The universal relations studied in \citep{Bretz:2015rna} were those among stellar multipole moments rather than the I-Love-Q relations.} has been studied in \citep{Bretz:2015rna} within the Newtonian limit and small differential-rotation approximation. The authors showed that the fractional difference in the relations from the uniformly-rotating case is comparable to the fractional amount of differential rotation over uniform rotation. On the other hand, the effect of finite temperature (as well as the composition dependence) has been studied in \citep{Martinon:2014uua}. The authors showed that when proto-NSs are formed, the I-Love-Q relations are different from the original ones for cold NSs by up to 20\%, which is much larger than the EoS-variation within the relations. However, several seconds after their births, the relations reduce to the original ones. Similarly, just after formation of a NS, either through single star evolution or in neutron star mergers, deviations from beta equilibrium may persist for several seconds. During this time, the deviation of the neutron to proton ratio from the beta equilibrium value may again act as an additional parameter in the equation of state. This effect has not been explored yet and is left for future work.


\section*{Acknowledgements}
We would like to thank Kuantay Boshkayev for helping us reproduce some of the results in previous literature.
K.Y. acknowledges support from NSF Award PHY-1806776, a Sloan Foundation Research Fellowship and the Ed Owens Fund. 
K.Y. would like to also acknowledge support by the COST Action GWverse CA16104 and JSPS KAKENHI Grants No. JP17H06358.

\bibliography{mybib}

\begin{thebibliography}{}
\makeatletter
\relax
\def\mn@urlcharsother{\let\do\@makeother \do\$\do\&\do\#\do\^\do\_\do\%\do\~}
\def\mn@doi{\begingroup\mn@urlcharsother \@ifnextchar [ {\mn@doi@}
  {\mn@doi@[]}}
\def\mn@doi@[#1]#2{\def\@tempa{#1}\ifx\@tempa\@empty \href
  {http://dx.doi.org/#2} {doi:#2}\else \href {http://dx.doi.org/#2} {#1}\fi
  \endgroup}
\def\mn@eprint#1#2{\mn@eprint@#1:#2::\@nil}
\def\mn@eprint@arXiv#1{\href {http://arxiv.org/abs/#1} {{\tt arXiv:#1}}}
\def\mn@eprint@dblp#1{\href {http://dblp.uni-trier.de/rec/bibtex/#1.xml}
  {dblp:#1}}
\def\mn@eprint@#1:#2:#3:#4\@nil{\def\@tempa {#1}\def\@tempb {#2}\def\@tempc
  {#3}\ifx \@tempc \@empty \let \@tempc \@tempb \let \@tempb \@tempa \fi \ifx
  \@tempb \@empty \def\@tempb {arXiv}\fi \@ifundefined
  {mn@eprint@\@tempb}{\@tempb:\@tempc}{\expandafter \expandafter \csname
  mn@eprint@\@tempb\endcsname \expandafter{\@tempc}}}

\bibitem[\protect\citeauthoryear{{Abbott} et~al.,}{{Abbott}
  et~al.}{2018}]{GW170817}
{Abbott} B.~P.,  et~al., 2018, \mn@doi [\prl] {10.1103/PhysRevLett.121.161101},
  \href {https://ui.adsabs.harvard.edu/abs/2018PhRvL.121p1101A} {121, 161101}

\bibitem[\protect\citeauthoryear{{Benacquista}}{{Benacquista}}{2011}]{benacquista}
{Benacquista} M.~J.,  2011, \mn@doi [\apj] {10.1088/2041-8205/740/2/L54}, \href
  {https://ui.adsabs.harvard.edu/abs/2011ApJ...740L..54B} {740, L54}

\bibitem[\protect\citeauthoryear{{Boshkayev} \& {Quevedo}}{{Boshkayev} \&
  {Quevedo}}{2018}]{hotwd}
{Boshkayev} K.,  {Quevedo} H.,  2018, \mn@doi [\mnras] {10.1093/mnras/sty1227},
  \href {http://adsabs.harvard.edu/abs/2018MNRAS.478.1893B} {478, 1893}

\bibitem[\protect\citeauthoryear{{Boshkayev}, {Quevedo}, {Kalymova}  \&
  {Zhami}}{{Boshkayev} et~al.}{2014}]{hartle}
{Boshkayev} K.,  {Quevedo} H.,  {Kalymova} Z.,   {Zhami} B.,  2014, arXiv
  e-prints, \href {http://adsabs.harvard.edu/abs/2014arXiv1409.2472B} {}

\bibitem[\protect\citeauthoryear{{Boshkayev}, {Quevedo}  \&
  {Zhami}}{{Boshkayev} et~al.}{2017}]{ilq_wd}
{Boshkayev} K.,  {Quevedo} H.,   {Zhami} B.,  2017, \mn@doi [\mnras]
  {10.1093/mnras/stw2614}, \href
  {http://adsabs.harvard.edu/abs/2017MNRAS.464.4349B} {464, 4349}

\bibitem[\protect\citeauthoryear{Bretz, Yagi  \& Yunes}{Bretz
  et~al.}{2015}]{Bretz:2015rna}
Bretz J.,  Yagi K.,   Yunes N.,  2015, \mn@doi [Phys. Rev.]
  {10.1103/PhysRevD.92.083009}, D92, 083009

\bibitem[\protect\citeauthoryear{{Chatziioannou}, {Haster}  \&
  {Zimmerman}}{{Chatziioannou} et~al.}{2018}]{measuring_NS_MR_using_ILQ}
{Chatziioannou} K.,  {Haster} C.-J.,   {Zimmerman} A.,  2018, \mn@doi [\prd]
  {10.1103/PhysRevD.97.104036}, \href
  {https://ui.adsabs.harvard.edu/abs/2018PhRvD..97j4036C} {97, 104036}

\bibitem[\protect\citeauthoryear{{Flanagan} \& {Hinderer}}{{Flanagan} \&
  {Hinderer}}{2008}]{ns_conservative_effect}
{Flanagan} {\'E}.~{\'E}.,  {Hinderer} T.,  2008, \mn@doi [Physical Review D]
  {10.1103/PhysRevD.77.021502}, \href
  {https://ui.adsabs.harvard.edu/abs/2008PhRvD..77b1502F} {77, 021502}

\bibitem[\protect\citeauthoryear{{Hartle}}{{Hartle}}{1967}]{realhartle}
{Hartle} J.~B.,  1967, \mn@doi [\apj] {10.1086/149400}, \href
  {https://ui.adsabs.harvard.edu/abs/1967ApJ...150.1005H} {150, 1005}

\bibitem[\protect\citeauthoryear{{Hartle} \& {Thorne}}{{Hartle} \&
  {Thorne}}{1968}]{1968ApJ...153..807H}
{Hartle} J.~B.,  {Thorne} K.~S.,  1968, \mn@doi [\apj] {10.1086/149707}, \href
  {https://ui.adsabs.harvard.edu/abs/1968ApJ...153..807H} {153, 807}

\bibitem[\protect\citeauthoryear{{Iben}, {Tutukov}  \& {Fedorova}}{{Iben}
  et~al.}{1998}]{tidal_heating}
{Iben} Icko J.,  {Tutukov} A.~V.,   {Fedorova} A. r.~V.,  1998, \mn@doi [The
  Astrophysical Journal] {10.1086/305972}, \href
  {https://ui.adsabs.harvard.edu/abs/1998ApJ...503..344I} {503, 344}

\bibitem[\protect\citeauthoryear{{Komatsu}, {Eriguchi}  \& {Hachisu}}{{Komatsu}
  et~al.}{1989}]{form_of_omega}
{Komatsu} H.,  {Eriguchi} Y.,   {Hachisu} I.,  1989, \mn@doi [\mnras]
  {10.1093/mnras/237.2.355}, \href
  {https://ui.adsabs.harvard.edu/abs/1989MNRAS.237..355K} {237, 355}

\bibitem[\protect\citeauthoryear{{Littenberg}}{{Littenberg}}{2011}]{many_wds_lisa}
{Littenberg} T.~B.,  2011, \mn@doi [\prd] {10.1103/PhysRevD.84.063009}, \href
  {https://ui.adsabs.harvard.edu/abs/2011PhRvD..84f3009L} {84, 063009}

\bibitem[\protect\citeauthoryear{Martinon, Maselli, Gualtieri  \&
  Ferrari}{Martinon et~al.}{2014}]{Martinon:2014uua}
Martinon G.,  Maselli A.,  Gualtieri L.,   Ferrari V.,  2014, \mn@doi [Phys.
  Rev.] {10.1103/PhysRevD.90.064026}, D90, 064026

\bibitem[\protect\citeauthoryear{{Mora} \& {Will}}{{Mora} \&
  {Will}}{2004}]{love_nums}
{Mora} T.,  {Will} C.~M.,  2004, \mn@doi [\prd] {10.1103/PhysRevD.69.104021},
  \href {https://ui.adsabs.harvard.edu/abs/2004PhRvD..69j4021M} {69, 104021}

\bibitem[\protect\citeauthoryear{Passamonti, Stavridis  \& Kokkotas}{Passamonti
  et~al.}{2008}]{Passamonti:2007td}
Passamonti A.,  Stavridis A.,   Kokkotas K.,  2008, \mn@doi [Phys. Rev.]
  {10.1103/PhysRevD.77.024029}, D77, 024029

\bibitem[\protect\citeauthoryear{{Paxton} et~al.,}{{Paxton}
  et~al.}{2013}]{mesa}
{Paxton} B.,  et~al., 2013, \mn@doi [\apjs] {10.1088/0067-0049/208/1/4}, \href
  {http://adsabs.harvard.edu/abs/2013ApJS..208....4P} {208, 4}

\bibitem[\protect\citeauthoryear{{Pelisoli}, {Kepler}, {Koester},
  {Castanheira}, {Romero}  \& {Fraga}}{{Pelisoli} et~al.}{2018}]{lowmass_wd}
{Pelisoli} I.,  {Kepler} S.~O.,  {Koester} D.,  {Castanheira} B.~G.,  {Romero}
  A.~D.,   {Fraga} L.,  2018, \mn@doi [Monthly Notices of the Royal
  Astronomical Society] {10.1093/mnras/sty1101}, \href
  {https://ui.adsabs.harvard.edu/abs/2018MNRAS.478..867P} {478, 867}

\bibitem[\protect\citeauthoryear{{Piro}}{{Piro}}{2011}]{piro}
{Piro} A.~L.,  2011, \mn@doi [\apj] {10.1088/2041-8205/740/2/L53}, \href
  {https://ui.adsabs.harvard.edu/abs/2011ApJ...740L..53P} {740, L53}

\bibitem[\protect\citeauthoryear{{Piro}}{{Piro}}{2019}]{new_piro}
{Piro} A.~L.,  2019, \mn@doi [\apjl] {10.3847/2041-8213/ab44c4}, \href
  {https://ui.adsabs.harvard.edu/abs/2019ApJ...885L...2P} {885, L2}

\bibitem[\protect\citeauthoryear{Poisson}{Poisson}{1998}]{Poisson:1997ha}
Poisson E.,  1998, \mn@doi [Phys. Rev.] {10.1103/PhysRevD.57.5287}, D57, 5287

\bibitem[\protect\citeauthoryear{{Salpeter}}{{Salpeter}}{1961}]{1961ApJ...134..669S}
{Salpeter} E.~E.,  1961, \mn@doi [\apj] {10.1086/147194}, \href
  {https://ui.adsabs.harvard.edu/abs/1961ApJ...134..669S} {134, 669}

\bibitem[\protect\citeauthoryear{{Shah} \& {Nelemans}}{{Shah} \&
  {Nelemans}}{2014}]{shah_and_nelemans2}
{Shah} S.,  {Nelemans} G.,  2014, \mn@doi [\apj] {10.1088/0004-637X/790/2/161},
  \href {https://ui.adsabs.harvard.edu/abs/2014ApJ...790..161S} {790, 161}

\bibitem[\protect\citeauthoryear{{Shah}, {van der Sluys}  \& {Nelemans}}{{Shah}
  et~al.}{2012}]{shah_and_nelemans1}
{Shah} S.,  {van der Sluys} M.,   {Nelemans} G.,  2012, \mn@doi [\aap]
  {10.1051/0004-6361/201219309}, \href
  {https://ui.adsabs.harvard.edu/abs/2012A%26A...544A.153S} {544, A153}

\bibitem[\protect\citeauthoryear{{Shapiro} \& {Teukolsky}}{{Shapiro} \&
  {Teukolsky}}{1986}]{book}
{Shapiro} S.~L.,  {Teukolsky} S.~A.,  1986, {Black Holes, White Dwarfs and
  Neutron Stars: The Physics of Compact Objects}

\bibitem[\protect\citeauthoryear{Stavridis, Passamonti  \& Kokkotas}{Stavridis
  et~al.}{2007}]{Stavridis:2007xz}
Stavridis A.,  Passamonti A.,   Kokkotas K.,  2007, \mn@doi [Phys. Rev.]
  {10.1103/PhysRevD.75.064019}, D75, 064019

\bibitem[\protect\citeauthoryear{{Yagi} \& {Yunes}}{{Yagi} \&
  {Yunes}}{2013a}]{ilq_og}
{Yagi} K.,  {Yunes} N.,  2013a, \mn@doi [\prd] {10.1103/PhysRevD.88.023009},
  \href {http://adsabs.harvard.edu/abs/2013PhRvD..88b3009Y} {88, 023009}

\bibitem[\protect\citeauthoryear{{Yagi} \& {Yunes}}{{Yagi} \&
  {Yunes}}{2013b}]{second_ILQ_paper}
{Yagi} K.,  {Yunes} N.,  2013b, \mn@doi [Science] {10.1126/science.1236462},
  \href {https://ui.adsabs.harvard.edu/abs/2013Sci...341..365Y} {341, 365}

\bibitem[\protect\citeauthoryear{{Yagi} \& {Yunes}}{{Yagi} \&
  {Yunes}}{2016}]{binary_love_relations}
{Yagi} K.,  {Yunes} N.,  2016, \mn@doi [Classical and Quantum Gravity]
  {10.1088/0264-9381/33/13/13LT01}, \href
  {https://ui.adsabs.harvard.edu/abs/2016CQGra..33mLT01Y} {33, 13LT01}

\bibitem[\protect\citeauthoryear{{Yagi} \& {Yunes}}{{Yagi} \&
  {Yunes}}{2017}]{approx_universal_relations}
{Yagi} K.,  {Yunes} N.,  2017, \mn@doi [Classical and Quantum Gravity]
  {10.1088/1361-6382/34/1/015006}, \href
  {https://ui.adsabs.harvard.edu/abs/2017CQGra..34a5006Y} {34, 015006}

\bibitem[\protect\citeauthoryear{{Yagi}, {Stein}, {Pappas}, {Yunes}  \&
  {Apostolatos}}{{Yagi} et~al.}{2014}]{why_ILQ}
{Yagi} K.,  {Stein} L.~C.,  {Pappas} G.,  {Yunes} N.,   {Apostolatos} T.~A.,
  2014, \mn@doi [\prd] {10.1103/PhysRevD.90.063010}, \href
  {https://ui.adsabs.harvard.edu/abs/2014PhRvD..90f3010Y} {90, 063010}

\makeatother
\end{thebibliography}
\bibliographystyle{mnras}

\appendix
\section{Equations of Structure for Differentially-Rotating WDs} \label{sec:derive_eqns}
Previous works \citep{realhartle,1968ApJ...153..807H, hartle} have shown how the equations of structure of a rotating Newtonian configuration may be derived using the Hartle-Thorne formalism. These works assume that the effects of rotation are small and work perturbatively. In this section, we will briefly go over how the equations derived in \cite{realhartle} and \cite{hartle} are altered to account for differential rotation within Newtonian gravity (see \cite{Stavridis:2007xz,Passamonti:2007td} for a similar framework for relativistic stars). See Section \ref{sec:parametrized} for discussion on how we implement differential rotation.

\subsection{Background and Perturbed Equations}
We begin by writing down the equation of hydrostatic balance and the gravitational Poisson equation.
The gravitational potential $\Phi$ separates into its background contribution (order $\Omega^0$) and leading-order perturbations (order $\Omega^2$), additionally selecting out the $\ell = 0$ and $\ell = 2$ spherical harmonic modes of the perturbation. Thus, each equation becomes three separate equations.
%
The equation of hydrostatic equilibrium becomes
\begin{alignat}{2}
    \label{omega0}
    \Omega_c^0:& \hspace{1.5cm} \int \frac{dP}{\rho} + \Phi^{(0)} &&= \text{const},\\
    \label{omega2_l0}
    \Omega_c^2, \ell=0:& \hspace{0.4cm}\xi_0 \frac{d\Phi^{(0)}}{dr} + \Phi^{(2)}_0 + U_0 &&= \text{const}^{(2)}, \\
    \label{omega2_l2}
    \Omega_c^2, \ell=2:& \hspace{0.4cm}\xi_2 \frac{d\Phi^{(0)}}{dr} + \Phi^{(2)}_2 + U_2 &&= 0,
\end{alignat}
where $\xi$ is the perturbation to the radial coordinate $r \rightarrow r + \xi + \mathcal{O}(\Omega_c^4)$. Here, we have denoted the order in $\Omega_c$ by a superscript and the spherical harmonic by a subscript. The expansion of $U$ in spherical harmonics is not as simple as in the constant rotation case, so we calculate the $\ell = 0$ and $\ell = 2$ components here:
\begin{align}
    \label{U0}
    U_0(r) &= -\frac{1}{2} \Omega_c^2 r^2 f_0(\alpha), \\
    \label{U2}
    U_2(r) &= -\frac{1}{2} \Omega_c^2 r^2 f_2(\alpha),
\end{align}
where $\alpha \equiv r/A$, and the functions $f_0(\alpha)$ and $f_2(\alpha)$ are given by
\begin{align}
    f_0(\alpha) &= \frac{1}{2} \int_0^\pi \frac{\sin^2\theta}{1+\alpha^2\sin^2\theta} P_0(\cos{\theta}) \sin{\theta} d\theta \hspace{1cm} \nn \\
    \label{f0}
    &= \frac{1}{\alpha^3} \left( \alpha - \frac{\text{Arcsinh}(\alpha)}{\sqrt{1+\alpha^2}}\right), \\ \nn \\
    f_2(\alpha) &= \frac{5}{4} \int_0^\pi \frac{\sin^2\theta}{1+\alpha^2\sin^2\theta} P_2(\cos{\theta}) \sin{\theta} d\theta \hspace{1cm} \nn \\
    \label{f2}
    &= \frac{5}{2\alpha^5} \left(3\alpha - \frac{(3+2\alpha^2) \text{Arcsinh}(\alpha)}{\sqrt{1+\alpha^2}}\right).
\end{align}
The gravitational Poisson equation, which states
\beq
\label{poisson}
\grad^2 \Phi = 4\pi G \rho,
\eeq
becomes
\begin{alignat}{2}
    \label{poisson_background}
    \Omega_c^0:& \hspace{2.35cm} \grad^2_r \Phi^{(0)} &&= 4\pi G \rho, \\
    \Omega_c^2, \ell = 0:& \hspace{0.4cm} \xi_0 \frac{d}{dr} \grad^2_r \Phi^{(0)} + \grad^2_r \Phi^{(2)}_0 &&= 0, \\
    \Omega_c^2, \ell = 2:& \hspace{0.4cm} \xi_2 \frac{d}{dr} \grad^2_r \Phi^{(0)} + \grad^2_r \Phi^{(2)}_r - \frac{6}{r^2} \Phi^{(2)}_2 &&= 0.
\end{alignat}

We now define two new variables $p_0^*$ and $m_0^*$ to simplify the above equations:
\begin{align}
    \label{p0star_def}
    p_0^* &\equiv \xi_0 \frac{d\Phi^{(0)}}{dr}, \\
    \label{m0star_def}
    \frac{G m_0^*}{r^2} &\equiv \frac{d\Phi^{(2)}_0}{dr}.
\end{align}
It can be shown that, when integrated from the center to the surface, $m_0^*$ is the correction to the mass. See \cite{hartle} for a more explicit discussion of these new variables (our $m_0^*$ is their $M^{(2)}$). 

These six equations plus the definition of interior mass
\beq
\frac{dm}{dr} = 4\pi r^2 \rho
\eeq
and the EoS (which is well-known for WDs) are all that are necessary to solve for the interior structure of a differentially-rotating WD. The stellar mass for a non-rotating configuration $M$ is determined from $M=m(R)$ with the stellar radius $R$ determined by the condition $P(R)=0$. 

Let us rewrite the above equations here for completeness. The background equations are:
\begin{align}
    \label{P_diffeq}
    \frac{dP}{dr} &= -\frac{G m \rho}{r^2}, \\
    \label{m_diffeq}
    \frac{dm}{dr} &= 4\pi r^2 \rho, \\
    \label{eos}
    P &= P(\rho).
\end{align}
The $\ell = 0$ equations are:
\begin{align}
    \label{p0star_diffeq}
    \frac{dp_0^*}{dr} &= - \frac{G m_0^*}{r^2} + \frac{1}{2} \Omega_c^2 \left(2rf_0(\alpha) + r^2 \frac{df_0(\alpha)}{dr}\right), \\
    \label{m0star_diffeq}
    \frac{dm_0^*}{dr} &= 4\pi r^2 \frac{d\rho}{dP}p_0^* \rho.
\end{align}
The $\ell = 2$ equations are a second-order ODE in terms of $\Phi^{(2)}_2$, which may be decomposed into two first-order ODEs to be numerically integrated:
\begin{align}
    \label{phi22_diffeq}
    \frac{d\Phi^{(2)}_2}{dr} &\equiv g^{(2)}_2, \\
    \label{g22_diffeq}
    \frac{dg^{(2)}_2}{dr} &= -4\pi G \rho \frac{d\rho}{dP} \left(\Phi^{(2)}_2 - \frac{1}{2}\Omega_c^2 r^2 f_2(\alpha)\right) + \frac{6}{r^2}\Phi^{(2)}_2 - \frac{2}{r} g^{(2)}_2.
\end{align}
We had three equations from decomposing both Equations \ref{hse_original} and \ref{poisson}, and we added in the definition of mass and the EoS, totalling eight equations, yet here we only have six (the two $\ell = 2$ equations are really just one equation). What happened to the seventh and eighth equations? The ``unused" equations are Equation \ref{omega2_l2} and \ref{poisson_background}, which we may use to solve for $\xi_2$ and  $\Phi^{(0)}$.

In the main part of this paper, we only kept to leading-order in spin, so the only perturbed variable we need is $\Phi^{(2)}_2$, which is used to calculate the quadrupole moment $Q$. However, in the interest of being thorough, we include the equations for the $\ell = 0$ variables (which tell us information about corrections to the mass, radius, and moment of inertia) as well.

\subsection{Boundary Conditions} \label{sec:boundary_conditions}
It is important to have knowledge of how the above functions behave at small $r$ away from the center of the star to have accurate initial conditions. For the background variables $\rho$, $m$, and $P$ near $r = 0$:
\begin{align}
    \rho(r) &= \rho_c \hspace{1cm} (\text{free parameter}),\\
    m(r) &= \frac{4}{3} \pi r^3 \rho_c, \\
    P(r) &= P(\rho_c),
\end{align}
where the central density $\rho_c$ is a free parameter to be chosen. 

For the $\ell = 0$ equations, we look at the leading-order terms in $r$ in Equation \ref{p0star_diffeq}. It is not immediately clear what the leading-order in $r$ is for the first term, but the term in parenthesis is clearly of order $r$ (see Equation \ref{U0}). We assume that this is the lowest order in $r$ for $p_0^*$. This would imply that $p_0^* \sim r^2$ near the center, which implies $m_0^* \sim r^5$ near the center. Plugging this back into Equation \ref{p0star_diffeq} confirms that linear-order is the lowest-order in $r$. Thus, the initial conditions are
\begin{align}
    p_0^*(r) &= \frac{1}{3} \Omega_c^2 r^2, \\
    m_0^*(r) &= \frac{4}{15}\pi \rho \frac{d\rho}{dP} \Omega_c^2 r^5.
\end{align}

For the $\ell = 2$ equations, we have two ODEs, so we clearly need two boundary conditions. The first is found by noting that, near the center, $\Phi^{(2)}_\ell \propto r^{\ell}$ to keep the solution finite at $r = 0$. The second boundary condition comes by matching the values obtained from the interior with the values obtained from the exterior at the surface of the star. In the exterior region of the star, the potential $\Phi^{(2)}_\ell \propto r^{-\ell-1}$ in order to keep the solution finite at infinity. Now, because the differential equations are linear in $\Phi^{(2)}_2$ and $g^{(2)}_2$, we can say that
\begin{align}
    \label{truephi_ab}
    \Phi^{(2)}_2(r) &= a(r) + b(r) \Phi(0), \\
    \label{trueg_cd}
    g^{(2)}_2(r) &= c(r) + d(r) \Phi(0),
\end{align}
where $a, b, c,$ and $d$ are some arbitrary functions of radius, and $\Phi(0)$ is assumed to be the true value of $\Phi^{(2)}$ near the center of the star. Then, we demand that at the surface
\beq
\frac{d\Phi^{(2)}_2}{dr}\Bigg|_{r = R} = -\frac{\ell + 1}{R} \Phi^{(2)}_2\Big|_{r = R} = -\frac{3}{R} \Phi^{(2)}_2\Big|_{r = R},
\eeq
which tells us that 
\beq
c(R) + d(R) \Phi(0) = -\frac{3}{R} \Big(a(R) + b(R) \Phi(0)\Big).
\eeq
Then, if $a, b, c$, and $d$ are known functions of $r$, we may then solve for the true initial condition $\Phi(0)$
\beq
\Phi(0) = -\frac{c(R) + 3 a(R)/R}{d(R) + 3 b(R)/R},
\eeq
which may then be substituted into Equations \ref{truephi_ab} and \ref{trueg_cd} to find the true functions $\Phi^{(2)}_2(r)$ and $g^{(2)}_2(r)$. Now all that remains is to find $a, b, c$, and $d$ as functions of radius. This is done by carefully choosing two values of $\Phi(0)$ and integrating the differential equations twice. The functions $a(r)$ and $c(r)$ are given by the results of an integration when $\Phi(0)$ is chosen to equal zero. Similarly, $b(r)$ and $d(r)$ are given by the results of an integration when $\Phi(0)$ is chosen to equal one and the functions $a(r)$ and $c(r)$ are subtracted off.

\subsection{Calculating Additional Variables} \label{sec:other_vars}
We now seek to calculate the remaining variables necessary to test the I-Love-Q relations. In the main body of this paper, we only kept up to leading order in spin, so that the mass, radius, and moment of inertia were entirely background quantities; additionally, we considered the angular momentum proportional to $\Omega_c$ and not to contain higher-order terms. However, in the interest of being thorough, we derive how the corrections to $I$ and $J$ may be calculated under this parametrized formulation of differential rotation. 

We begin by calculating the moment of inertia, $I$, which is given by the integral
\beq
I = \int dI = \int \rho (r\sin{\theta})^2 dV.
\eeq
It can be shown that to next-to-leading order in the perturbation, this integral becomes (see \citealt{hartle})
\beq
I = I^{(0)} + I^{(2)} = \frac{8\pi}{3} \int_0^{R} \rho(r) r^4 dr - \frac{8\pi}{3} \int_0^{R} \frac{d\rho}{dr} r^4 \left(\xi_0 - \frac{1}{5} \xi_2\right)dr,
\eeq
where $\xi_2$ is found via solving Equation \ref{omega2_l2}
\beq
\xi_2(r) = -\frac{r^2}{G m(r)} \Big(\Phi^{(2)}_2(r) - \frac{1}{2}\Omega_c^2 r^2 f_2(\alpha)\Big),
\eeq
and $d\rho/dr$ is found via the chain rule and the EoS
\beq
\frac{d\rho}{dr} = \frac{d\rho}{dP} \frac{dP}{dr} = -\left(\frac{dP}{d\rho}\right)^{-1} \frac{G m(r) \rho(r)}{r^2}.
\eeq

We now wish to calculate the total angular momentum of the star $J$. In constantly-rotating stars, this is simply equal to $I\Omega$, but we have now promoted $\Omega$ to be a function of radius. Thus, it is now absorbed into the volume integral, and we find
\begin{align}
    J^{(1)} &= \int \rho r^2 \sin^2\Theta \, \Omega_c \frac{A^2}{A^2 + r^2 \sin^2\theta} dV \\
    &= 2\pi \Omega_c \int_0^{R} dr \int_0^\pi \rho r^4 dr \frac{\sin^3\theta}{1+ \alpha^2 \sin^2\theta} d\theta \\
    &= 4\pi \Omega_c \int_0^R \rho r^4 f_0(\alpha) dr,
\end{align}
where we remind the reader that we have defined $\alpha \equiv r/A$. We have shown that $I$ may be split into a background piece and a perturbed piece of order $\Omega_c^2$. The same can be shown for $J$. What we have calculated above is the leading-order term in $J$, which is of order $\Omega_c^1$, hence the superscript (1). The term corresponding to the contribution from $I^{(2)}$ can be shown to be
\beq
J^{(3)} = -4\pi \Omega_c \int_0^R \frac{d\rho}{dr} r^4 \Big(\xi_0 f_0(\alpha) + \frac{1}{5}\xi_2 f_2(\alpha)\Big) dr.
\eeq
In the main body of this paper, only the terms $I^{(0)}$ and $J^{(1)}$ were used. 

Next, the rotational Love number may either be calculated using the Clairaut equation (see \cite{hartle}) or by simply calculating the ratio of the response potential to the forcing potential. The apsidal motion constant $k_2$ is given by
\beq
\label{defk2}
k_2 = \frac{1}{2} \frac{\Phi^{(2)}_2}{U_2}\Bigg|_R,
\eeq
and the Love number is simply $\lambda = 2 k_2$. In this work, we use the latter method to calculate Love numbers. Other works vary in their definition of the Love number (some define $\lambda$ to have units -- see e.g. \citealt{ilq_wd}); here, we define the rotational Love number as the response in the gravitational potential to the forcing centrifugal potential, making $\lambda$ unitless. There appears to be general agreement on the meaning of the apsidal motion constant $k_2$, so we note that our definition of the Love number is related by a factor of two to $k_2$.

Finally, we wish to calculate the quadrupole moment $Q$. The gravitational potential exterior to the WD is given by
\beq
\label{phitot1}
\Phi(r, \theta) = -\frac{GM}{r} + \frac{GQ}{r^3} P_2(\cos{\theta}).
\eeq
In how we have defined $m_0^*$ (see Equation \ref{m0star_def}), one can see that Equation \ref{phitot1} may be solved for $Q$:
\beq
\label{defQ}
\Phi^{(2)}_2\Big|_R = \frac{G Q}{R^3} \hspace{1cm} \rightarrow \hspace{1cm} Q = \frac{R^3}{G} \Phi^{(2)}_2\Big|_R.
\eeq
Using this sign convention, $Q > 0$ represents an oblate object, and $Q < 0$ represents a prolate object.

\section{Equilibrium Fluid Configurations in WDs} \label{sec:l0_l2_eqns}
In this section, we will derive formulae for the moment of inertia $I$ and the quadrupole moment $Q$ of a perturbed fluid configuration. In contrast to the Hartle-Thorne formalism, here we assume that the mass (rather than the central density) is fixed after the perturbation is ``turned on". 

We begin by assuming that the background (unperturbed) quantities are known as functions of radius: pressure $P$, density $\rho$, sound-speed squared $c_s^2$, Brunt-Vaisala frequency $N^2$, and interior mass $m$. Now, due to some perturbing potential $U$, the fluid configuration experiences small changes in these quantities away from their background values. In general, for some arbitrary fluid variable $B$, the Eulerian perturbation $\delta B$ is defined by
\beq
\delta B \equiv B(\mbf{x}, t) - B_0(\mbf{x}, t),
\eeq
where $B_0$ is the background quantity. See \cite{book}, Section 6.2 for a more thorough description of these perturbations. 

Next, we express all relevant perturbed quantities in terms of spherical harmonics:
\beq
\delta A = \sum_{\ell, m} \delta A_{\ell m}(r) Y_{\ell m}(\theta, \phi), 
\eeq
with $A = (P,\rho,\Phi,\xi_r,\xi_h,U)$. 
Here, $\xi_r$ is the radial perturbation, $\xi_h$ is the horizontal perturbation, and $\Phi$ is the gravitational potential. One can show (e.g. \citealt{book}) that to conserve the mass of the fluid element, the following relation must hold:
\beq
\label{cons_mass}
\delta \rho = -\vgrad \cdot \Big(\rho \mbf{\xi}\Big).
\eeq

We define the quadrupole moment via the gravitational potential $\Phi$ of the fluid:
\beq
\Phi(\mbf{x}) = -\frac{GM}{r} - \frac{GQ}{r^3} P_2(\cos{\theta}) + \mathcal{O}\left(\frac{R^4}{r^5}\right),
\eeq
where $G$ is the gravitational constant, $M$ is the total mass of the fluid, $r$ is the distance from the origin to the point at which the gravitational potential is being evaluated, and $P_2(x)$ is the $\ell=2$ Legendre polynomial in $x$. One can then show that $Q$ is given by
\beq
\label{defQ_fluid}
Q = \sqrt{\frac{4\pi}{5}} \int \delta \rho_{20}(r') (r')^4 dr',
\eeq
where $\rho_{20}$ is the $(\ell,m) = (2,0)$ mode of the density perturbation, using the language of spherical harmonics, rather than Legendre polynomials. 

Next, we wish to find the perturbation to the integral quantity $I$. In the absence of any perturbations, the moment of inertia is given by
\beq
I = \int_0^R \rho (x^2 + y^2) d^3\mbf{x}.
\eeq
Following the procedure of \cite{book}, the perturbation to $I$ using the Lagrangian treatment is given by
\begin{align}
    dI &= \int \Delta (x^2 + y^2) \times \rho d^3 \mbf{x} \\
    &= \int \rho \Big(2x \xi_x + 2y \xi_y\Big) d^3\mbf{x} \\
    &= 2 \int \rho \mbf{\xi} \cdot \Big(x \mbf{\Hat{x}} + y \mbf{\Hat{y}}\Big) d^3\mbf{x},
\end{align}
where $\Delta$ represents a Lagrange perturbation. To simplify the above dot product, we rewrite the radial perturbation as the sum of its radial and horizontal piece:
\beq
\mbf{\xi} = \xi_{r, \ell m} Y_{\ell m} \mbf{\Hat{r}} + \xi_{h, \ell m} r \vgrad Y_{\ell m}.
\eeq
One can show that the vector sum of $\mbf{x} + \mbf{y}$ can be expressed as
\beq
\mbf{x} + \mbf{y} = r\sin{\theta}\Big(\sin{\theta} \mbf{\Hat{r}} + \cos{\theta}\mbf{\Hat{\theta}}\Big).
\eeq
Then, using the orthogonality of the gradient of spherical harmonics,
\beq
\int \Big(r\vgrad Y_{\ell m}\Big) \cdot \Big(r \vgrad Y_{\ell' m'}\Big) d\Omega = \ell (\ell + 1) \delta_{\ell \ell'} \delta_{m m'},
\eeq
one can show that the perturbation to $I$ is given by
\begin{align}
    \label{delta_I}
    dI = 2\int \rho r^3 dr \Bigg[\frac{4}{3}\sqrt{\pi} \Big(\xi_{r,0}(r) - \frac{1}{\sqrt{5}}\xi_{r,2}(r)\Big) - 4\sqrt{\frac{\pi}{5}}\xi_{h,2}(r)\Bigg],
\end{align}
where only the $\ell = 0, 2$ and $m = 0$ components survive the integration. Here, we have dropped the $m$ subscript, as it is zero for all terms. Thus, we need to know $\xi_{r, \ell}$ and $\xi_{h, \ell}$. 

Now, for a fluid configuration exposed to some perturbing potential (with no oscillatory response), the equation of hydrostatic balance becomes (to leading-order in the perturbation)
\beq
\label{perturbed_hse}
0 = - \vgrad \delta P - \vgrad \Big(\delta \rho \, \Phi + \rho \, \delta \Phi\Big) - \rho \vgrad U.
\eeq
The gradient operator acts both on the radial piece in the spherical harmonic expansion as well as the spherical harmonics themselves. Thus, we may split Equation \ref{perturbed_hse} into a radial equation and a horizontal equation. Each term in the radial equation carries a spherical harmonic, which we may cancel from each. Similarly, the horizontal expression carries the gradient of a spherical harmonic, which is proportional to $\ell/r$. We keep the $\ell/r$ and cancel the rest, leaving us with:
\begin{align}
    \label{hse_radial}
    0 &= -\frac{d \delta P_{\ell m}}{dr} - g \delta \rho_{\ell m} - \rho \Big( \frac{d \delta \Phi}{dr} + \frac{dU_{\ell m}}{dr} \Big), \\
    \label{hse_horiz}
    0 &= \frac{\ell}{r} \Big[ \delta P + \rho \big(\delta \Phi + U\Big)\Big].
\end{align}
For $\ell = 0$, the second equation tells us no information. Thus, we must solve the equations of structure separately for the $\ell = 0$ and $\ell = 2$ cases. 

\subsection{Solving the $\ell = 2$ case}
We begin with the simpler $\ell = 2$ case. We assume that all background quantities ($P$, $\rho$, $m(r)$, $N^2$, $c^2$) are known as functions of radius. Then, we have the three equations we discussed above (Equations \ref{cons_mass}, \ref{hse_radial}, and \ref{hse_horiz}), as well as the EoS:
\beq
\delta \rho = \frac{\delta P}{c^2} + \rho \frac{N^2}{g} \xi_r,
\eeq
where $g$ is the interior gravity (equal to $Gm(r)/r^2$), and we have cancelled the $Y_{\ell m}$ from both sides and suppressed the $\ell m$ subscript. For the rest of this section, the $\ell m$ subscripts are implied on all perturbed quantities unless specifically stated otherwise. From the horizontal hydrostatic equilibrium equation, we have that
\beq
\frac{\delta P}{\rho} = -\Big(\delta \Phi + U\Big).
\eeq
We then substitute this into the EoS to find
\beq
\delta \rho = -\frac{\rho}{c^2} \Big(\delta \Phi + U\Big) + \rho \frac{N^2}{g} \xi_r,
\eeq
which we may then plug into the radial hydrostatic equilibrium equation. Some cancellation occurs, and we are left with
\beq
0 = -\frac{N^2}{g}\rho - \rho N^2 \xi_r.
\eeq
In the above simplification, we have used the definition of $N^2$:
\beq
N^2 \equiv -g \left(\frac{1}{\rho} \frac{d\rho}{dr} + \frac{g}{c^2} \right).
\eeq
Thus, in radiative regions (where $N^2$ > 0), we have that
\beq
\label{solve_xir_l2}
\xi_r  = - \frac{\delta \Phi + U}{g}.
\eeq
In convective regions, where $N^2 \lesssim 0$, this relation does not hold explicitly. Then, we may solve for $\delta \rho$ to yield
\beq
\label{delta_rho}
\delta \rho = - \frac{d\rho}{dr}\xi_r.
\eeq
We now need to invoke a fourth equation: the gravitational Poisson equation:
\begin{align}
\grad^2 \delta \Phi &= 4\pi G \delta \rho \\
&= 4\pi G \left(\frac{d\rho}{dr}\right) \left(\frac{\delta \Phi + U}{g}\right) \\
&= 4\pi G \rho \left( \frac{1}{c^2} + \frac{N^2}{g^2}\right)\Big(\delta\Phi + U\Big).
\end{align}
This is a self-consistent equation in terms of $\delta \Phi$, its derivatives, and known quantities. Thus, with the proper boundary conditions, this may be solved to find $\delta \Phi$, which tells us $\xi_r$ and then $\delta \rho$ and $\delta P$. Once all of these variables are known, we substitute back into the mass-conservation equation to solve for $\xi_h$:
\beq
\label{xih}
\xi_h = \frac{r}{\ell(\ell+1)} \left(\frac{\delta \rho}{\rho} + \frac{d\xi_r}{dr} + \frac{2\xi_r}{r} + \frac{1}{\rho}\frac{d\rho}{dr}\xi_r\right).
\eeq
The radial derivative of $\xi_r$ may be found by taking the derivative of Equation \ref{solve_xir_l2} and carefully applying the chain rule. Since $U$ is a known quantity, all derivatives of terms in Equation \ref{solve_xir_l2} are known explicitly. All we need now are the proper boundary conditions for $\delta \Phi$. We refer the reader to the discussion in Appendix \ref{sec:boundary_conditions} on choosing appropriate boundary conditions for $\Phi$. The process is similar: two different initial conditions near $r = 0$ are chosen for $\delta \Phi$ in order to solve for the appropriate initial condition by matching with the values at the surface.

\subsection{Solving the $\ell = 0$ case}
Next, we turn our attention to the $\ell=0$ case. From Equation \ref{delta_I}, we see that we must also know the $\ell = 0$ mode of $\xi_r$ to properly integrate to find the perturbations to $I$. The derivation above relies on the horizontal hydrostatic equilibrium equation (Equation \ref{hse_horiz}), which only applies in the case where $\ell \neq 0$. To solve for the $\ell = 0$ case, we must start over, using only the radial hydrostatic equilibrium equation, the equation of mass-conservation, the EoS, and the gravitational Poisson equation. In the previous section, we also had the horizontal hydrostatic equilibrium equation. The variable $\xi_h$ is zero in the $\ell = 0$ mode (see Equation \ref{xih}) to keep the number of equations and variables equal. We define a new variable
\beq
\Psi \equiv \frac{\delta P}{\rho} + \delta \Phi + U,
\eeq
where as before, we have dropped the $\ell m$ subscripts (zero is assumed for both here). We rewrite the four equations we have in terms of this new variable, as well as for clarity:
\begin{align}
    0 &= -\frac{d}{dr}\Big(\rho(\Psi - \delta \Phi - U)\Big) - g\delta \rho - \rho \frac{d}{dr}\Big(\delta \Phi + U\Big), \\
    0 &= \delta \rho + \frac{d}{dr}\Big(\rho \xi_r\Big), \\
    \delta \rho &= \frac{\rho}{c^2} \Big(\Psi - \delta \Phi - U\Big) + \rho \frac{N^2}{g}\xi_r, \\
    \grad^2 \delta \Phi &= 4\pi G \delta \rho.
\end{align}
Substituting the EoS into the radial hydrostatic equilibrium equation and dividing by $\rho$, we find some cancellations, and we are left with
\beq
\frac{d\Psi}{dr} = N^2 \left( \frac{\Psi - \delta \Phi - U}{g} - \xi_r\right).
\eeq
For $\ell \neq 0$, we have defined $\Psi = 0$, so then also its derivative must be zero. This reduces to the equation for $\xi_r$ we found in Equation \ref{solve_xir_l2}. Next, we substitute $\delta \rho$ from the EoS into the equation of constant mass to arrive at:
\beq
\frac{d\xi_r}{dr} = -\xi_r \left(\frac{2}{r} - \frac{g}{c^2}\right) - \frac{\Psi - \delta \Phi - U}{c^2}.
\eeq
Finally, we substitute $\delta \rho$ from the EoS into the gravitational Poisson equation to arrive at a single equation containing derivatives of $\delta \Phi$. For numerical integration, it is best to rewrite this second-order ODE as two first-order ODEs, which we list below:
\begin{align}
    \frac{d\delta \Phi}{dr} &\equiv \delta g, \\
    \frac{d\delta g}{dr} &= 4\pi G \rho \left( \frac{\Psi - \delta \Phi - U}{c^2} + \frac{N^2}{g}\xi_r\right) - \frac{2}{r} \delta g.
\end{align}

We now have four first-order ODEs for our four variables. We therefore need four sets of boundary conditions in order to solve for these entirely. Two of these boundary conditions are the restrictions we have previously imposed on $\delta \Phi$ and $\delta g$:
\begin{align}
     \delta g(r) \Big|_{r = R} &= -\frac{1}{R} \delta \Phi(r) \Big|_{r = R}, \\
     \delta g(r) \Big|_{r \sim 0} &= 0.
\end{align}
The next boundary condition comes by demanding that the surface of both the rotating and non-rotating star have zero pressure, asserting that $\Delta P = 0$ (the Lagrangian perturbation to P; see \citealt{book}, Chapter 6). This gives us
\beq
\Psi(r)\Big|_{r = R} = \Big(g(r) \xi_r(r)\Big)\Big|_{r = R} + \delta \Phi(r) \Big|_{r = R} + U(r)\Big|_{r = R}.
\eeq
For the fourth bounday condition, we look at the behavior of $\xi_r(r)$ near the center. Just as we needed $\delta \Phi \propto r^\ell$ for it to be finite near the center, we can see that we need $\xi_r \propto r$ for it to be finite near the center as well. Thus, we have our fourth boundary condition:
\beq
\xi_r(r)\Big|_{r \sim 0} = -\frac{r}{3c^2} \Big(\Psi(r) - \delta \Phi(r)\Big) \Big|_{r\sim0}.
\eeq

We now have four ODEs for four variables with four boundary conditions on those variable, though these boundary conditions are given in terms of ``free" parameters: $\delta \Phi(0)$, $\delta \Phi(R)$, $\Psi(0)$, and $\Psi(R)$. How do we find the correct values of these free parameters? We integrate multiple times with different boundary conditions from the surface and from the center, matching our values at some radius in the middle. We begin by declaring (in a similar method to what we used in Appendix \ref{sec:boundary_conditions}) that the four variables we study here are exclusively determined by these ``free" parameters:
\begin{align}
    \delta \Phi(r) &= a_1(r) + a_2(r) \delta \Phi(0) + a_3(r) \Psi(0), \\
    \delta g(r) &= b_1(r) + b_2(r) \delta \Phi(0) + b_3(r) \Psi(0), \\
    \xi_r(r) &= c_1(r) + c_2(r) \delta \Phi(0) + c_3(r) \Psi(0), \\
    \Psi(r) &= d_1(r) + d_2(r) \delta \Phi(0) + d_3(r) \Psi(0).
\end{align}
We may rewrite these four equations in terms of a matrix (noting that $a_i$ through $d_i$ are all functions of $r$):
\begin{align}   \label{matrix_from_center}
    \left( {\begin{array}{c}
    \delta \Phi \\
    \delta g \\
    \xi_r \\
    \Psi
    \end{array} } \right) = \left( {\begin{array}{ccc}
    a_1 & a_2 & a_3 \\
    b_1 & b_2 & b_3 \\
    c_1 & c_2 & c_3 \\
    d_1 & d_2 & d_3 
    \end{array} } \right) \left( {\begin{array}{c}
    1 \\
    \delta \Phi(0) \\
    \Psi(0)
    \end{array} } \right).
\end{align}
We may do the same for the ``free" parameters at the surface:
\begin{align}   \label{matrix_from_surface}
    \left( {\begin{array}{c}
    \delta \Phi \\
    \delta g \\
    \xi_r \\
    \Psi
    \end{array} } \right) = \left( {\begin{array}{ccc}
    a_1' & a_2' & a_3' \\
    b_1' & b_2' & b_3' \\
    c_1' & c_2' & c_3' \\
    d_1' & d_2' & d_3' 
    \end{array} } \right) \left( {\begin{array}{c}
    1 \\
    \delta \Phi(R) \\
    \Psi(R)
    \end{array} } \right).
\end{align}
Now, these ``free" parameters are not truly free; they have true values that are not able to be chosen, and we wish to find these true values. If the functions $a_i$ through $d_i'$ are known functions of $r$, then we solve for these true values by matching at some radius in between the center and the surface, which we will call $r_1$. Then, at $r_1$, we demand that
\begin{align}
    \left( {\begin{array}{ccc}
    a_1 & a_2 & a_3 \\
    b_1 & b_2 & b_3 \\
    c_1 & c_2 & c_3 \\
    d_1 & d_2 & d_3 
    \end{array} } \right) \left( {\begin{array}{c}
    1 \\
    \delta \Phi(0) \\
    \Psi(0)
    \end{array} } \right) = 
    \left( {\begin{array}{ccc}
    a_1' & a_2' & a_3' \\
    b_1' & b_2' & b_3' \\
    c_1' & c_2' & c_3' \\
    d_1' & d_2' & d_3' 
    \end{array} } \right) \left( {\begin{array}{c}
    1 \\
    \delta \Phi(R) \\
    \Psi(R)
    \end{array} } \right).
\end{align}
We have four equations for four unknowns, which we may also express as a matrix (where all quantities other than the ``free" parameters are again evaluated at $r = r_1$):
\begin{align}   \label{true_inits}
    \left( {\begin{array}{c}
    a_1' - a_1 \\
    b_1' - b_1 \\
    c_1' - c_1 \\
    d_1' - d_1 
    \end{array} } \right) = \left( {\begin{array}{cccc}
    a_2 & -a_2' & a_3 & -a_3' \\
    b_2 & -b_2' & b_3 & -b_3' \\
    c_2 & -c_2' & c_3 & -c_3' \\
    d_2 & -d_2' & d_3 & -d_3'
    \end{array} } \right)\left( {\begin{array}{c}
    \delta \Phi(0) \\
    \delta \Phi(R) \\
    \Psi(0) \\
    \Psi(R)
    \end{array} } \right).
\end{align}
This matrix may be inverted, and thus the true values of the ``free" parameters are solved, once $a_i$ through $d_i'$ are known functions of $r$. 

How do we find these functions of $r$? We integrate several times with different choices of the boundary conditions. We choose to integrate three times from the surface and three times from the center. In both cases (surface and center) the choices of boundary conditions are the same. We begin by choosing the tuple of boundary conditions $(\delta \Phi, \Psi) = (0, 0)$. Then, the results of this integration give us
\beq
\delta \Phi_1(r) (\text{known}) = a_1(r),
\eeq
which gives us the coefficient $a_1$ (or $a_1'$) as a function of radius. This integration tells us all of the variables with subscript 1. For the next integration, we choose the boundary condition-tuple $(\delta \Phi, \Psi) = (0, 1)$, giving us
\beq
\delta \Phi_2(r) (\text{known}) = a_1(r) (\text{known}) + a_2(r),
\eeq
which then tells us $a_2$ (and similarly $a_2'$ and all other variables with subscript 2). Finally, we choose the third boundary condition-tuple $(\delta \Phi, \Psi) = (1, 0)$, which gives us
\beq
\delta \Phi_3(r) (\text{known}) = a_1(r) (\text{known}) + a_3(r),
\eeq
and we then know the rest of the coefficients $a_i$ through $d_i'$ as functions of radius. Thus, integrating the same set of differential equations three times (at both the center and the surface) gives us the coefficients, which we use to solve for the true boundary conditions via Equation \ref{true_inits}. Then, once the true boundary conditions are known, the true values of all four perturbed variables may be found using either Equation \ref{matrix_from_center} or \ref{matrix_from_surface}.

\bsp	
\label{lastpage}
\end{document}